\documentclass[aoas,preprint]{imsart}

\RequirePackage[OT1]{fontenc}
\RequirePackage{amsthm,amsmath,amsfonts,amssymb,bbm,bm}
\RequirePackage{natbib}
\usepackage{hyperref}
\usepackage{cleveref}
\usepackage{float}
\usepackage[linesnumbered, ruled]{algorithm2e}
\usepackage{multicol}
\usepackage{multirow}
\usepackage{textcomp}
\usepackage{caption}
\usepackage{subcaption}
\usepackage{booktabs}
\usepackage{graphicx} 
\usepackage{array}
\usepackage{xcolor}
\usepackage[normalem]{ulem}

\newcolumntype{P}[1]{>{\centering\arraybackslash}p{#1}}
\SetKwInput{KwData}{Inputs}
\SetKwInput{KwResult}{Output}

\newcommand{\Up}[1]{\mathrm{#1}}

\startlocaldefs
\numberwithin{equation}{section}
\endlocaldefs

\begin{document}

\begin{frontmatter}
\title{Markov Decision Processes with Dynamic Transition Probabilities: An Analysis of Shooting Strategies in Basketball}

\runtitle{Analyzing Shooting Strategies in Basketball}

\begin{aug}
\author{\fnms{Nathan} \snm{Sandholtz}\ead[label=e1]{nsandhol@sfu.ca}}
\and
\author{\fnms{Luke} \snm{Bornn}}

\runauthor{N. Sandholtz, L. Bornn}

\affiliation{Simon Fraser University}

\address{N. Sandholtz, L. Bornn\\
Department of Statistical and Actuarial Science\\
Simon Fraser University\\ 
8888 University Drive\\
Burnaby, BC, V5A 1S6\\
Canada\\
\printead{e1}\\
\phantom{E-mail:\ nsandhol@sfu.ca}}
\end{aug}

\begin{abstract}
In this paper we model basketball plays as episodes from team-specific non-stationary Markov decision processes (MDPs) with shot clock dependent transition probabilities.  Bayesian hierarchical models are employed in the modeling and parametrization of the transition probabilities to borrow strength across players and through time.  To enable computational feasibility, we combine lineup-specific MDPs into team-average MDPs using a novel transition weighting scheme.  Specifically, we derive the dynamics of the team-average process such that the expected transition count for an arbitrary state-pair is equal to the weighted sum of the expected counts of the separate lineup-specific MDPs. 

We then utilize these non-stationary MDPs in the creation of a basketball play simulator with uncertainty propagated via posterior samples of the model components.  After calibration, we simulate seasons both on-policy and under altered policies and explore the net changes in efficiency and production under the alternate policies.  Additionally, we discuss the game-theoretic ramifications of testing alternative decision policies. 
\end{abstract}


\end{frontmatter}

\section{Introduction} \label{sec:intro}

\begin{figure}[H]
\centering
\begin{subfigure}{.33\textwidth}
\begin{flushleft}
  \includegraphics[trim={.5cm 1cm .5cm 2.5cm},clip,width=1.85in]{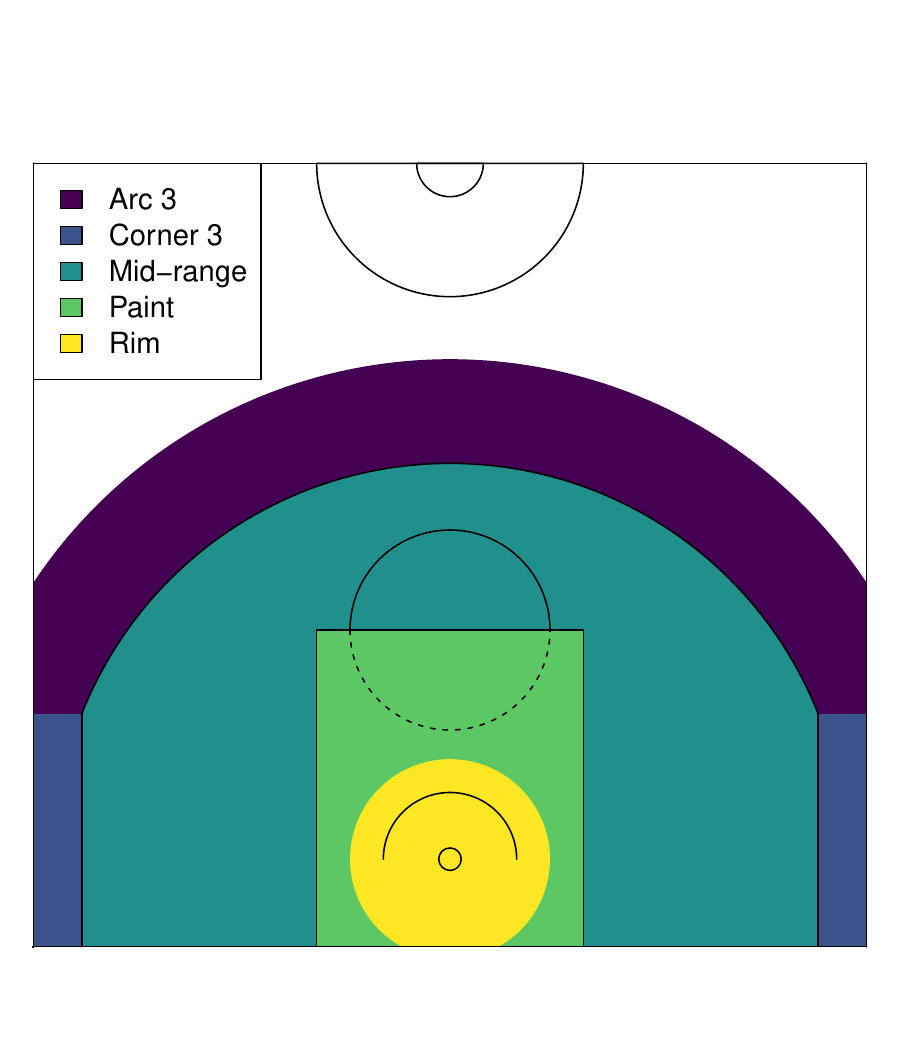}
  \caption{}
  \end{flushleft}
  \label{fig:a}
\end{subfigure}%
\begin{subfigure}{.67\textwidth}
\begin{flushright}
\includegraphics[width=3in]{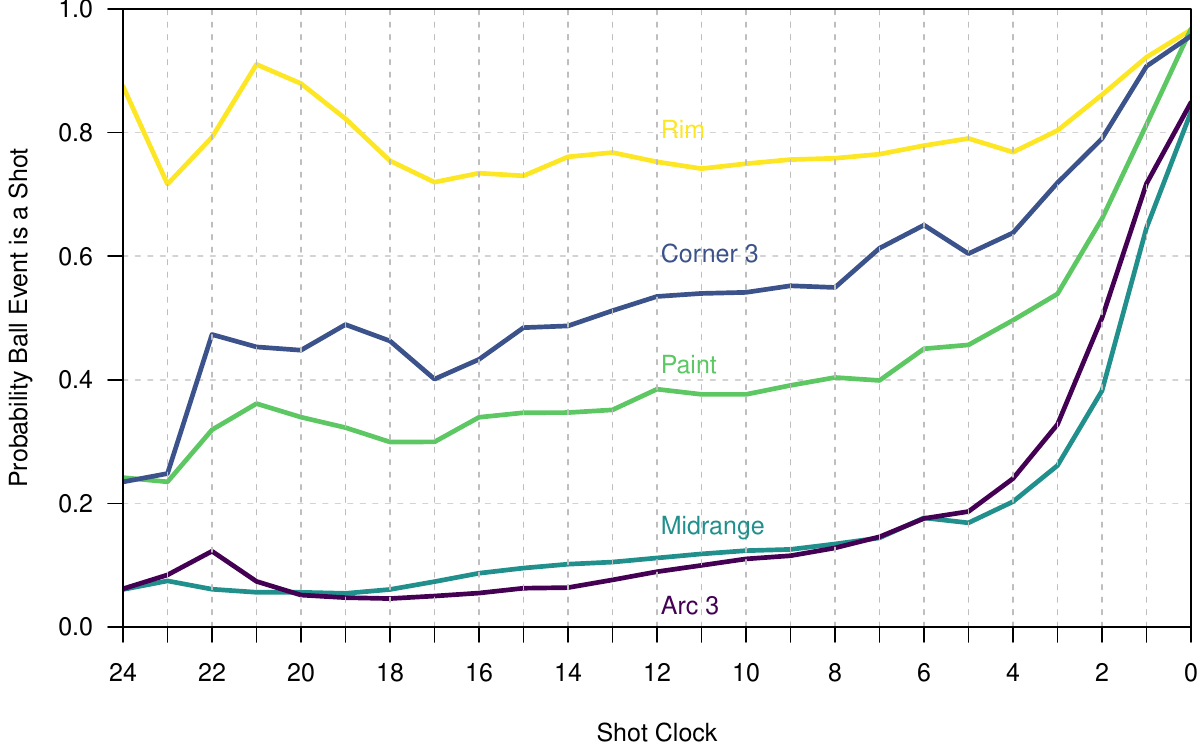} 
\caption{}
\end{flushright}
  \label{fig:b}
\end{subfigure}
\setlength{\belowcaptionskip}{-5pt}
\setlength{\abovecaptionskip}{-5pt}
\caption{(a) Breakdown of court locations as used in our models and simulations. Rim: within a 6 ft radius of the center of the hoop; Paint: outside the restricted area but within the key; Mid-range: outside of the paint but within the 3-point line; Corner 3: beyond the 3-point line but below the break of the 3-point arc; Arc 3: beyond the 3-point line, above the arc 3 break, but within 30 ft. of the hoop; Backcourt: all locations beyond the arc 3 region. (b) Empirical league-average shot policies for the 2015-16 NBA regular season.   We see lower shot probabilities in the mid-range and arc 3 regions because the on-ball events in these regions are dominated by passes and dribbles.}
\label{fig:intro_figure}
\end{figure}

A basketball game can be framed as a collection of episodes from complex stochastic processes.  Each episode, or play, is comprised of a finite number of transitions between players and locations ultimately terminating in a shot, turnover, or foul.  An integral attribute of the game is that it is non-stationary; the transition probabilities are not constant over the 24 seconds in which a team has to shoot the ball.  For example, consider the relationship between time on the shot clock, which counts down these 24 seconds, and the probability of taking a shot as shown in Figure \ref{fig:intro_figure}.  The plot in (\subref{fig:b}) shows empirical league-average shot policies, which we define as the probability that any on-ball event (i.e. dribbles, passes, and shots) will be a shot, for the set of court regions defined in (\subref{fig:a}).  As the shot clock winds down, the probability of shooting increases---quite dramatically in the final seconds of the shot clock.  

Determining optimal policies for player shooting is a critical problem in the game of basketball and it remains an active area of research \citep{skinner2015optimal, goldman2014optimal}.   However, the inherent non-stationarity introduced by the shot clock makes assessing shot selection optimality a complex problem.  In this project, we propose a method to test and compare shot policies which accounts for the dynamic nature of a basketball play.  Two critical assumptions underlying our approach are that shot policies are both time-varying and malleable.  Basketball analysts often focus on the less flexible component of shot efficiency---field goal percentage, or the percentage of a player's shots that he makes.  Improving shooting skill can take years of practice, whereas the shot policy is comparatively controllable; players \textit{choose} where and how often they shoot when they have the ball in their possession.  

Given the malleable nature of shot policies, we explore what could have happened if a player's shot policy had changed.  To enable this exploration, we model plays as episodes from latent Markov decision processes (MDPs) with dynamic within-episode transition probabilities.  We approximate these functional transition probabilities via transition probability tensors (TPTs), then estimate the latent components of the MDP using Bayesian hierarchical models.  Our method involves combining several Markov chains with overlapping state spaces into an average Markov chain, which we derive subject to the constraint that the expected total transition count for an arbitrary state-pair is equal to the weighted sum of the expected counts of the separate chains.  

We then develop a method to simulate from these processes not simply by outcome, but rather at the sub-second level, incorporating every intermediary and terminal on-ball event over the course of a play.  The uncertainty in our estimation of the MDP gets propagated into the simulations via posterior samples of the MDP model.  Ultimately, our method allows us to make distributional estimates of counter-factual scenarios such as, ``What could have happened if a team took contested mid-range shots less frequently early in the shot clock?"  While we focus primarily on shot policies in this paper, narrowing in on a player's choice to shoot or not at any given instant, the framework presented here can be altered to accommodate the whole space of decisions players can make on-ball, including movement and passing.

\subsection{Related work and contributions}

This paper adds to the growing literature of spatiotemporal analyses of team invasion sports (e.g. basketball, football, soccer, and hockey).  We refer the reader to \cite{gudmundsson2016} for a survey.  Within this body of work, Markov models have significant presence: \cite{Goldner2012} uses a Markov model as a framework for evaluating plays in American football; \cite{Hirotsu2002} use Markov processes to determine optimal substitution patterns in an English Premier League match; and \cite{thomas2013competing} use a semi-Markov process to model team scoring rates in hockey.

The landmark work of \cite{cervone2016multiresolution} is particularly relevant to the methods we introduce in this paper.  The state space and hierarchical models we develop have similarities to the coarsened possession process they employ.  However, our ultimate goals are fundamentally different.  \cite{cervone2016multiresolution} aim to estimate instantaneous point values of possessions whereas we utilize a decision process framework to estimate the macro-effects if player decisions were to change.  

We approach the problem similarly to \cite{Routley2015}, who applies a Markov game formalism to value player actions in hockey, incorporating context and a lookahead window in time.  As in \cite{Routley2015}, we do not aim to compute optimal strategies.  Instead, we provide a basketball play simulator by which alternate policies can be explored.  Since the defense is not an adversarial agent in our model but is built into the system via the probabilistic components of the MDP, this simulator is proposed as an exploratory tool as opposed to a mechanism to compute policy optima.  
  
Several papers have endeavored to simulate a basketball game using Markov models \citep{vstrumbelj2012simulating, vravcar2016}.  Our simulator is unique among these studies in a number of ways.  We account for the uncertainty in every estimated parameter, propagating this uncertainty through to the simulations.  Also, though \cite{vravcar2016} incorporate game-clock time, these simulation methods do not account for the inherent non-stationarity \textit{within} a possession introduced by the shot clock.  We propose a novel method to account for the non-stationarity of basketball plays using tensors in the MDP framework.  By incorporating this dependency in our model, we can explore far more detailed policy changes with correspondingly more accurate results, particularly with respect to shot clock violations and time-specific policy changes within plays.  

This work also contributes to the literature and practical application of discrete absorbing Markov chains.  We formalize a method to construct and estimate an average chain from several independent Markov chains with overlapping state spaces.  The term ``average'' as used here refers to a chain that yields the same number of state-to-state transition counts in expectation for all unique state pairs spanned by the set of independent chains.  This allows us to reliably estimate aggregate counts across the system of chains without having to estimate each chain individually.  This result is critical in making our method both parsimonious and computationally feasible. 

\subsection{Description of data}

We use high-resolution optical tracking data collected by STATS LLC from the 2015-2016 NBA regular season.  These data include the $x,y$ coordinates of all 10 players on the court and the $x,y,z$ coordinates of the ball at a frequency of 25 observations per second.  These data are further annotated with features such as shots, passes, dribbles, fouls, etc.  For this project we subset the data to observations with tagged ball-events including dribbles, passes, rebounds, turnovers, and shots.  This significantly reduces the number of observations while retaining the core structure of a possession.  We have created a companion GitHub repository for this paper, including a simplified walkthrough of the methods presented in this paper and one game of data provided by STATS LLC.\footnote{\url{https://github.com/nsandholtz/nba_replay}}

\subsection{Outline}

The rest of the paper is outlined as follows: In Section 2 we give a brief overview of Markov decision processes and describe how we incorporate tensors in the framework of an MDP.  We also detail the construction of an average chain from several independent Markov chains with overlapping state spaces.  In Section 3 we define probabilistic models for each latent component of the MDP.  The inference procedures used in fitting the models is detailed and we illustrate the model fits.  In Section 4 we describe how we simulate plays from team-specific MDPs and show calibration results from our simulations.  In Section 5 we show the results of our simulations under various altered policies and discuss potential game-theoretic ramifications of altering policies.  Our concluding remarks comprise Section 6.  

\section{Decision process framework}

We begin this section with a brief overview of Markov decision processes and frame the process in basketball terms.  Next, we show how to construct an average chain from several independent Markov chains with overlapping state spaces.  The section concludes with our framework for incorporating tensors into the transition dynamics and action policies in order to account for the nonstationarity of these probabilities in time.

\subsection{Markov decision processes} \label{sec:MDP}

Markov decision processes are utilized in many modern reinforcement learning problems to characterize interactions between an agent and his environment.  In this paper we restrict our attention to finite MDPs, which can be represented as a tuple: $\langle \mathcal{S},\mathcal{A},P(\cdot),R(\cdot)\rangle$, where $\mathcal{S}$ represents a discrete and finite set of states, $\mathcal{A}$ represents a finite set of actions the agent can take, $P(\cdot)$ defines the transition probabilities between states, and $R(\cdot)$ defines the immediate reward the agent receives for any given state/action pair.  The agent operates in the environment according to a policy, $\pi(\cdot)$, which defines the probabilities that govern the agent's choice of action based on the current state of the environment.  $\pi(\cdot)$ is the only aspect of the system that the agent controls.  Typically, the agent's goal is to maximize their long-term rewards by making iterative modifications to their policy based on the rewards they receive.  

We can define these functions succinctly in mathematical terms:
\begin{align}
P(s,a,s') &= \mathbb{P}[\Up{S}_{n+1} = s' ~ |~ \Up{S}_{n} = s, \Up{A}_{n} = a] \label{eq:T} \\ 
R(s,a) &= \mathbb{E} [\Up{R}_{n+1}  ~ |~ \Up{S}_{n} = s, \Up{A}_{n} = a] \label{eq:R} \\ 
\pi(s,a) &= \mathbb{P}[\Up{A}_{n} = a~ |~ \Up{S}_{n} = s] \label{eq:pi}.
\end{align} 
In (\ref{eq:T}-\ref{eq:pi}), non-italicized uppercase letters (e.g. $\Up{S}_n, \Up{A}_n, \text{and }\Up{R}_{n+1}$) represent random variables while italicized lowercase letters represent realized values of the corresponding random variables.  Specifically, $\Up{S}_n$ represents the agent's state in step $n$ of an episode from the MDP, $\Up{A}_n$ represents the action taken in that step, and $\Up{R}_{n+1}$ is the subsequent reward given that state-action pair.

For our basketball context, we define an episode as the sequence of events comprising a single play from start to finish.   A play is initialized in some state $s_0$, and $\pi(\cdot)$ determines the probability that the ballcarrier takes a shot (or other actions, as we explore later) given that state.  If he decides not to shoot, $P(\cdot)$ governs the probability of the ball entering any other state given the current state.  If he does take a shot, $R(\cdot)$ dictates the expected point value of that shot.  Figure A.1 in \ref{suppA} shows a graph of these relationships in context of a basketball play.  In our case, the governing probabilities are unobserved for each of these components and hence must be estimated.  We refer the reader to \cite{sutton2018reinforcement} and \cite{puterman2014markov} for a more expansive introduction to reinforcement learning and Markov decision processes.

\subsection{State and action space} \label{sec:state_and_action}
Following \cite{cervone2016multiresolution}, at any given moment, the state of a team's MDP is given by the identity of the ballcarrier, his court region, and an indicator of his defensive pressure (open or contested).  Formally, $\mathcal{S}^{team}$ is defined by the Cartesian product of three sets:
\begin{align}
\mathcal{S}^{team} = \mathcal{S}^{player} \times \mathcal{S}^{region}  \times \mathcal{S}^{defense}, \label{eq:cart_space} 
\end{align} 
where $\mathcal{S}^{player}$ is simply the set of all players on the team's roster, $\mathcal{S}^{region}$ comprises the six regions shown in Figure \ref{fig:intro_figure}(\subref{fig:b}), and defensive pressure is binary (`open' or `contested').  At any given moment $n$ of a play, $\Up{S}^{player}_n $ is determined by the ballcarrier, $\Up{S}^{region}_n$ is a simple function of his $x,y$ coordinates, and $\Up{S}^{defense}_n$ is based on both his nearest defender's distance (ndd) and his court region at that moment.\footnote{The specific rules we follow for determining contested vs. open shots are given by (A.4) in \ref{suppA}.}

To avoid these lengthy superscripts, going forward we will index players by $x \in \mathcal{S}^{player}$, court regions by $y \in \mathcal{S}^{region}$, and defensive pressure by $z \in \mathcal{S}^{defense}$.  Consequently, from this point onward, $x$ and $y$ no longer refer to Cartesian coordinates but rather to players and court regions respectively.  Putting this all together, $s^{(x,y,z)}_n$ denotes an observed state from $\mathcal{S}^{team}$  at the $n^{th}$ step of an episode from the team's MDP.

As we are primarily interested in shooting decisions, we have chosen a binary action space.  At each step in the process, the ballcarrier decides to either shoot or not shoot ($\mathcal{A} = \{\text{`Shoot', `Not Shoot'}\}$) according to his policy $\pi(\cdot)$.  If $a_n = $ `Shoot', the play terminates; otherwise, the subsequent transition is generated by $P(\cdot)$.  Later, we explore changes to passing probabilities via perturbations to  $P(\cdot)$.    

\subsection{Defining the average chain}
Because most teams use upwards of 500 lineups over the course of a season, we assume that $P(\cdot)$, $R(\cdot)$, and $\pi(\cdot)$ are invariant to the lineup, i.e., other players do not impact transitions and rewards.  This allows us to approximate the data generating process (i.e. state transitions and player actions) at any given point in time using a single team-average transition probability matrix, which is essential in developing both a tractable model for player-specific transitions and a computationally feasible method to simulate multiple season's worth of plays. We construct this team-average chain such that it yields the same number of state pair transitions in expectation as the sum across all the independent chains for every unique state pair spanned by the set of lineups.  Technically, for a fixed interval in time, a basketball play cannot yield an arbitrarily high number of transient transitions before absorption due to the process of the shot clock.  In this sense, the episodes we observe come from \textit{censored} Markov chains.   However, as this result could be useful in many Markov chain applications outside of basketball, we will detail this derivation in general limiting terms.

Consider a process defined by iterative episodes from two absorbing Markov chains, $\mathbf{M}^1$ and $\mathbf{M}^2$ (with initial distributions $\mathbf{d}^1$ and $\mathbf{d}^2$), where each successive episode is randomly determined to come from $\mathbf{M}^1$ or $\mathbf{M}^2$ with weights $w^1$ and $w^2 = (1 - w^1)$ respectively.\footnote{There are many terms throughout this paper that must be indexed across multiple dimensions.  For example, in this derivation, we need to differentiate one transition probability matrix $\mathbf{M}$ from another while also being able to index the $(j,k)$ elements of these separate matrices.  To communicate indexes over various dimensions, we have chosen to use superscripts in addition to subscripts to differentiate these dimensions that must be indexed.  For this reason, superscripts do not refer to exponents throughout this paper with the exception of inverses (e.g. $\mathbf{X}^{-1}$), and variance parameters, which are exclusively denoted by $\sigma^2$.}  Each chain, indexed by $\ell  \in \{1, 2\}$, has transient states $\mathcal{T}^{\ell} = \{t^{\ell}_1,\ldots, t^{\ell}_T\}$ and absorbing states $\mathcal{A}^{\ell} = \{a^{\ell}_1, \ldots, a^{\ell}_A\}$.  For simplicity, we assume that each chain has the same total number of transient states and absorbing states (i.e. $|\mathcal{T}^{\ell}| = T \text{ and } |\mathcal{A}^{\ell}| = A$ for each $\ell \in \{1, 2\}$), however, we note that set $\mathcal{T}^1$ is not equal to set $\mathcal{T}^2$, though there may exist overlapping states.\footnote{$\mathcal{A}^{\ell}, a^{\ell}$, and $A$ as defined here are not the same quantities defined previously in Sections \ref{sec:MDP} and \ref{sec:state_and_action}.  We felt that this abuse of notation was warranted in order to make the derivation clearer to the reader.}  Our task is to construct a chain, $\mathbf{M}^{\text{AVG}}$, such that the expected transition count for an arbitrary state-pair from one episode of this chain is equal to the weighted average of the expected transition counts of the separate chains for this same state-pair.  

We first write $\mathbf{M}^1$ and $\mathbf{M}^2$ in canonical form:

\begin{multicols}{2}
\begin{center}

$
\begin{array}{c c} &
\begin{array}{c c} ~ \mathcal{T}^1 & \mathcal{A}^1~\\
\end{array} 
\\
\mathbf{M}^1 =~~
\begin{array}{r}
\mathcal{T}^1 \\
\mathcal{A}^1 
\end{array} 
&
\left[
\begin{array}{c | c}
\mathbf{Q}^1  & \mathbf{U}^1   \\
\hline
\mathbf{0}  & \mathbf{I}   \\
\end{array}
\right] 
\end{array}
$ 

\end{center} 
\columnbreak
 \begin{center}

$
\begin{array}{c c} &
\begin{array}{c c} ~ \mathcal{T}^2 & \mathcal{A}^2~\\
\end{array} 
\\
\mathbf{M}^2 =~~
\begin{array}{r}
\mathcal{T}^2 \\
\mathcal{A}^2 
\end{array} 
&
\left[
\begin{array}{c | c}
\mathbf{Q}^2  & \mathbf{U}^2   \\
\hline
\mathbf{0}  & \mathbf{I}   \\
\end{array}
\right] 
\end{array}
$ 

\end{center}
\end{multicols}
\noindent where $\mathbf{I}$  is an $A$-by-$A$ identity matrix, $\mathbf{0}$ is an $A$-by-$T$ zero matrix, each $\mathbf{U}^{\ell}$ is a nonzero $T$-by-$A$ matrix containing the absorption probabilities for chain $\ell$, and each $\mathbf{Q}^{\ell}$ is a $T$-by-$T$ matrix of the transient state to transient state probabilities.  Each individual $(j,k)$ entry of $\mathbf{M}^{\ell}$ is the probability of immediately transitioning to state $t^{\ell}_k$ (or $a^{\ell}_k$) given current state $t^{\ell}_j$.

Following conventional notation (e.g. \cite{grimstead1997introduction}), we define the \textit{fundamental matrix} for chain $\mathbf{M}^{\ell}$ as $\mathbf{N}^{\ell} = (\mathbf{I} - \mathbf{Q}^{\ell})^{-1}$.  The $n^{\ell}_{ij}$ entry of $\mathbf{N}^{\ell}$ is the expected number of times that the chain visits transient state $t^{\ell}_j$ given that the episode is initialized in state $t^{\ell}_i$.   Next, we define the matrix of expected state-pair transition counts, $\mathbf{S}^{\ell}$, in a cell-wise manner such that the $(j,k)$ cell of $\mathbf{S}^{\ell}$ equals
\begin{align}
s^{\ell}_{jk} =  \sum_i d^{\ell}_i n^{\ell}_{ij} m^{\ell}_{jk}, \label{eq:s_l}
\end{align}
where $\mathbf{d}^{\ell} = \{d^{\ell}_1, \ldots, d^{\ell}_{T}\}$ is the initial distribution of $\mathbf{M}^{\ell}$.  In (\ref{eq:s_l}), $i \in \{1,\ldots, T\}$ indexes the initial starting state of the chain, $j \in \{1,\ldots, T\}$ indexes the origin state, and $k \in \{1,\ldots, T+A\}$ indexes the destination state.  Hence $\mathbf{S}^{\ell}$ is a $T$-by-$(T+A)$ matrix.

Conceptually, to create the average chain $\mathbf{M}^{\text{AVG}}$, we will combine $\mathbf{S}^1$ and $\mathbf{S}^2$ proportional to the number of episodes that come from each chain using their respective weights, then normalize the rows of the resulting matrix.  We then define matrix $\mathbf{S}^{\text{AVG}}$ such that any entry of this matrix equals the weighted average of the expected transition counts of the separate chains.  This requires a bit more notation.  For each chain $\mathbf{M}^{\ell}$, we define set $\mathcal{V}^{\ell}$ as the outer product of its transient state space $\mathcal{T}^{\ell}$ with its total state space $\mathcal{T}^{\ell} \cup \mathcal{A}^{\ell}$.  Specifically, $\mathcal{V}^{\ell} = \mathcal{T}^{\ell} \times \{\mathcal{T}^{\ell} \cup \mathcal{A}^{\ell}\} = \{(t^{\ell}_1,t^{\ell}_1), (t^{\ell}_1,t^{\ell}_2), (t^{\ell}_2,t^{\ell}_1), \ldots \}$.  $\mathbf{S}^{\text{AVG}}$ can then be defined cell-wise as  
\begin{align}
s^{\text{\tiny{AVG}}}_{j'k'} =  
\begin{cases} 
     w^1s^{1}_{j'k'} + (1 - w^1)s^{2}_{j'k'}, & (t_{j'}, t_{k'}) \in \mathcal{V}^1 \cap \mathcal{V}^2 \\
     w^1s^1_{j'k'}, & (t_{j'}, t_{k'}), \in \mathcal{V}^1 \cap (\mathcal{V}^2)^c \\
      (1 - w^1)s^2_{j'k'}, & (t_{j'}, t_{k'}) \in (\mathcal{V}^1)^c \cap \mathcal{V}^2 \\
      0, & \text{otherwise},
   \end{cases} \label{eq:s_avg}
\end{align}
where $j' \in \{1,\ldots, |\mathcal{T}^1 \cup \mathcal{T}^2|\}$ indexes the collective set of origin states (i.e. $\mathcal{T}^1 \cup \mathcal{T}^2$) and $k' \in \{1,\ldots, |(\mathcal{T}^1 \cup \mathcal{T}^2) \cup (\mathcal{A}^1 \cup \mathcal{A}^2)|\}$ indexes the collective set of destination states.  Therefore, $\mathbf{S}^{\text{AVG}}$ is a $|(\mathcal{T}^1 \cup \mathcal{T}^2)|$-by-$|(\mathcal{T}^1 \cup \mathcal{T}^2) \cup (\mathcal{A}^1 \cup \mathcal{A}^2)|$ matrix.

Finally, let 
\begin{center}
$
\begin{array}{c c} &
\begin{array}{c c} ~ \mathcal{T}^1 \cup \mathcal{T}^2 & \mathcal{A}^1 \cup \mathcal{A}^2~\\
\end{array} 
\\
\mathbf{M}^{\text{AVG}} =~~
\begin{array}{r}
 \mathcal{T}^1 \cup \mathcal{T}^2 \\
\mathcal{A}^1 \cup \mathcal{A}^2
\end{array} 
&
\left[
\begin{array}{c | c}
\mathbf{Q}^{\text{AVG}}  & \mathbf{U}^{\text{AVG}}   \\
\hline
\mathbf{0}  & \mathbf{I}^{|\mathcal{A}^1 \cup \mathcal{A}^2|}   \\
\end{array}
\right] 
\end{array}
$ 
\end{center}
where $\mathbf{I}$  is an $|\mathcal{A}^1 \cup \mathcal{A}^2|$-by-$|\mathcal{A}^1 \cup \mathcal{A}^2|$ identity matrix and the $(j',k')$ cells of $\mathbf{Q}^{\text{AVG}}$ and $\mathbf{U}^{\text{AVG}}$ are defined 
\begin{align}
q^{\text{\tiny{AVG}}}_{j'k'} &= \frac{s^{\text{\tiny{AVG}}}_{j'k'}}{\sum_i s^{\text{\tiny{AVG}}}_{j'i}}  \label{eq:q_avg} \\
u^{\text{\tiny{AVG}}}_{j'k'} &= \frac{s^{\text{\tiny{AVG}}}_{(j')(k'+|\mathcal{T}^1 \cup \mathcal{T}^2|)}}{\sum_i s^{\text{\tiny{AVG}}}_{j'i}}. \label{eq:u_avg}
\end{align} 
In (\ref{eq:q_avg})-(\ref{eq:u_avg}), $i \in \{1, \ldots, |(\mathcal{T}^1 \cup \mathcal{T}^2) \cup (\mathcal{A}^1 \cup \mathcal{A}^2)|\}$ and $j'$ and $k'$ are indexed as in (\ref{eq:s_avg}).  Clearly the rows of this matrix sum to 1 and all entries are non-negative, thereby making it a valid TPM.  By (\ref{eq:s_avg}), it is trivial that the expected transition count for an arbitrary state-pair from $\mathbf{M}^{\text{AVG}}$ is equal to the weighted average of the $\mathbf{M}^1$ and $\mathbf{M}^2$ expected state-pair transition counts.  Finally, the $j'$th element of the initial distribution, $\mathbf{d}^{\text{AVG}}$, for $\mathbf{M}^{\text{AVG}}$ is 
\begin{align}
d^{\text{\tiny{AVG}}}_{j'} =  
\begin{cases} 
     w^1d^{1}_{j'} + (1 - w^1)d^{2}_{j'}, & t_{j'} \in \mathcal{T}^1 \cap \mathcal{T}^2 \\
     w^1d^1_{j'}, & t_{j'}, \in \mathcal{T}^1 \cap (\mathcal{T}^2)^c \\
      (1 - w^1)d^2_{j'}, & t_{j'} \in (\mathcal{T}^1)^c \cap \mathcal{T}^2
   \end{cases} \label{eq:d_avg}
\end{align}
where $j' \in \{1,\ldots, |\mathcal{T}^1 \cup \mathcal{T}^2|\}$.  While we have shown this only for two chains, we can make this same argument recursively with the current iteration of $\mathbf{M}^{\text{AVG}}$ and a subsequent chain to incorporate into the average, $\mathbf{M}^{n+1}$, showing that this result holds for an arbitrary number of Markov chains.

The average chain allows us to accurately estimate aggregate counts (e.g. over the course of the regular season) across all lineups without having to estimate each lineup's transition probabilities individually, making the problem tractable while still retaining enough detail to explore the nuanced questions we are interested in.  

\subsection{Transition and policy tensors} \label{sec:tensor_framework}

In most MDP applications the transition dynamics, $P(\cdot)$, are treated as being static while $R(\cdot)$ is assumed to vary temporally.  However, in this paper we assume the opposite; only the reward function is time-independent.  The reason for this is that in our case, time---or the shot clock rather---resets with each new episode of the process as opposed to continuing globally across episodes.  We are concerned about \textit{within-episode} temporal dynamics, whereas most MDP applications consider time globally.  As such, the way we consider non-stationarity is quite different than how it is primarily treated in the literature.  Our framework requires a functional form of $P(\cdot)$ and $\pi(\cdot)$, whereas these are conventionally modeled statically.\footnote{In reinforcement learning applications, $\pi(\cdot)$ typically gets updated as the agent learns more about his environment.  In this sense $\pi(\cdot)$ is dynamic, but this is different than the within-episode functional form for $\pi(\cdot)$ we refer to here.}  

To incorporate within-episode non-stationarity in $P(\cdot)$ and $\pi(\cdot)$, we propose using tensors to allow for dynamic transition probabilities and shot policies over the shot clock.  In the stochastic processes literature, the term `transition probability tensor' arises (albeit infrequently) in reference to the series of $m$ transition probability matrices induced by an $m^{th}$-order Markov chain (e.g. \cite{li2014}).  This is not what we mean by this term.  Rather, we refer to a transition (or policy) tensor as an approximation to a dynamic transition probability \textit{function} of a continuous temporal covariate, which in this case is the shot clock.  

To this end, we approximate $P(\cdot)$ and $\pi(\cdot)$ as tensors with $n_{\scriptscriptstyle{\text{TPT}}}$ matrix slices, each representing a transition probability matrix (TPM) for a $\frac{24}{n_{\scriptscriptstyle{\text{TPT}}}}$-second interval of the shot clock.  We want to choose $n_{\scriptscriptstyle{\text{TPT}}}$ to be as small as possible while still allowing our model to adequately describe the non-stationary dynamics over the 24-second shot clock.  From a model parsimony and computational feasibility point of view, minimizing the number of TPT slices is desirable; each additional TPT slice adds roughly one million more model parameters, which leads to a significant increase to the computational burden when fitting the model.  We settled on $n_{\scriptscriptstyle{\text{TPT}}}= 8$, meaning that each TPT slice summarizes the transition dynamics over a three-second interval of the shot clock.

\begin{figure}[ht]
\begin{center}
\includegraphics[width=5in,  trim={.3cm .3cm .3cm .3cm}, clip]{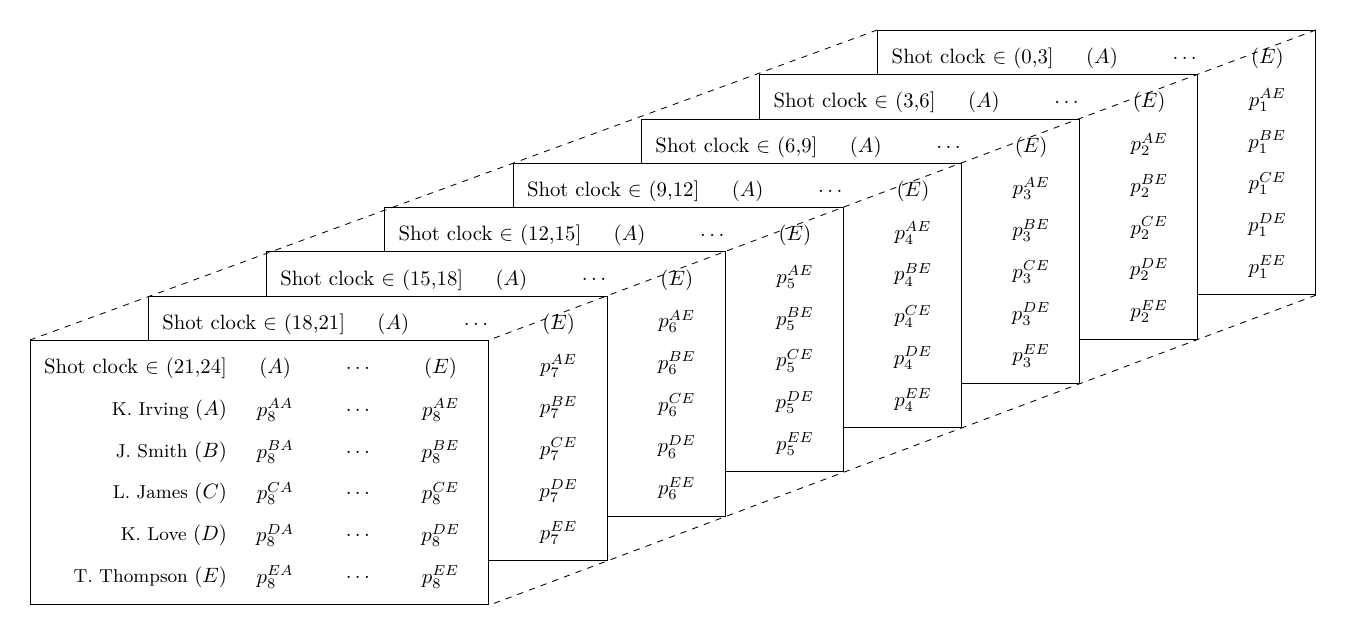} 
\end{center}
\setlength{\belowcaptionskip}{0pt}
\setlength{\abovecaptionskip}{0pt}
\caption{A concept illustration of $P(\cdot)$ for the Cleveland Cavaliers most common starting lineup in the 2015-16 NBA regular season.  For illustration purposes the row and column space of the TPT has been condensed to five single-player states, however, in our models a typical team has a state space of over 200 states.  Each slice represents an approximation of the state-to-state transition probabilities during a 3-second interval of the shot clock.  Hence in this example, $p^{AE}_8$  represents the average probability of Kyrie Irving passing the ball to Tristan Thompson when the shot clock is in the interval (21,24].}
\label{fig:tpm_3d}
\end{figure}

We effectuate this approximation via a simple indexing function $T(c_n)$:
\begin{align}
T(c_n) =  
\begin{cases} 
     1, & c_n \in (0, 3] \\
      & \vdots \\
     8, & c_n \in (21, 24] 
   \end{cases} \label{eq:slice_func}
\end{align}
where $c_n$ represents the shot clock time at the $n^{th}$ moment of a play.  Going forward we will represent realized values of $T(c_n)$ as $t_n$.  An illustrative 8-slice TPT is shown in Figure \ref{fig:tpm_3d} for the Cleveland Cavaliers 2015-16 most common starting lineup. 

The policy tensor is virtually identical to the transition probability tensor (TPT) in form, the only difference being the column space.  Since the ballcarrier makes only a binary decision at every step of the process, the shot policy given any time interval $t_n$, is a matrix with row space equal to that of the corresponding TPT slice and a column space of length two (`Shoot' and `Not shoot').  This tensor framework is the key to accurately exploring the effects of altering shot policies.  The efficiency of a shot is dependent on the time remaining on the shot clock, and this model framework allows us to account for this temporal dependency and tailor our policy alterations accordingly.  

\section{Hierarchical modeling and inference}
\label{sec:tensor_model_spec}

In this section, we propose models for the latent components of the MDP---$P(\cdot), R(\cdot)$, and $\pi(\cdot)$---with temporally varying dynamics as outlined in Section \ref{sec:tensor_framework}.  For each component we employ a Bayesian hierarchical modeling approach, which provides a natural way to share strength across parameters that are alike.  While we fit each model independently of the others, they each employ a common hierarchical structure---player-specific parameters borrow strength from position-specific parameters (e.g. point guards, power forwards, etc.), which in turn borrow strength from global location and defensive pressure parameters.

\subsection{Shot policy model---$\pi(\cdot)$} \label{sec:policy_model}
For an arbitrary play from a given team and letting $n_{\scriptscriptstyle{\text{TPT}}}= 8$, we model the probability that the $n^{th}$ action of the play is a shot as a function of the ballcarrier's propensity to take a shot given the current state of the MDP (defined by the ballcarrier, his court region, and defensive pressure), and the shot clock time interval:
\begin{align}
\pi(s, a) = \mathbb{P}(\Up{A}_n = \text{`Shoot'} | s^{(x,y,z)}_n, t_n, \boldsymbol \theta) &= \text{expit}(\theta^{(x,y,z)}_{t_n}).
 \end{align}
In this equation, $\Up{A}_n$ is a Bernoulli random variable for whether the $n^{th}$ action of a play is a shot,  $s^{(x,y,z)}_n$ is as defined in Section \ref{sec:state_and_action}, $t_n$ represents the interval of the shot clock at the $n^{th}$ moment of the play, $\theta^{(x,y,z)}_{t_n}$ denotes player $x$'s propensity to shoot the ball when in court region $y$ under defensive pressure $z$, and $\text{expit}(\cdot)$ is the inverse logit function: $\frac{\text{exp}(\cdot)}{1 + \text{exp}(\cdot)}$.  Note that $\boldsymbol \theta$ is a large parameter matrix of dimension $|\mathcal{S}^{team}|$-by-$n_{\scriptscriptstyle{\text{TPT}}}$.  For a 15-player roster using 3-second shot clock intervals for the TPT, this results in a 180-by-8 matrix.\footnote{Due to the extreme infrequency of  backcourt shots we don't estimate player-specific coefficients for backcourt shot policies and field goal percentages.  For notational simplicity we have omitted this technicality in the model definition.}  

We employ multi-level hierarchical priors for $\boldsymbol \theta$ to pool information across similar states of the MDP: 
\begin{align}
\boldsymbol \theta^{(x,y,z)} &\sim \mathbf{N}_8(\boldsymbol \beta^{(G(x),y,z)},  \boldsymbol \Sigma_{\theta}), \label{eq:policy_player} \\ 
\boldsymbol \beta^{(g,y,z)} &\sim \mathbf{N}_8(\boldsymbol \gamma^{(y,z)} ,\boldsymbol \Sigma_{\beta}), \label{eq:policy_position} \\
\boldsymbol \gamma^{(y,z)} &\sim \mathbf{N}_8(\mathbf{0}, \boldsymbol \Sigma_{\gamma}),  \label{eq:policy_location} 
\end{align}
where $G(x)$ returns the position type $g$ of player $x$ (e.g. point-guard, center, etc.).  In (\ref{eq:policy_player}), the 8-dimensional state-specific shot propensity parameter $\boldsymbol \theta^{(x,y,z)}$ is given a multivariate normal prior with mean vector $\boldsymbol \beta^{(G(x),y,z)}$, denoting the average shot propensity parameter vector for all players who have the same position as player $x$ (i.e. all players $\{x'_1, \ldots, x'_{N_g}\}$ for whom $G(x'_i) = G(x)$).  This effectively shrinks player $x$'s estimated shooting propensity toward players who share his same position and the shrinkage is more pronounced if he has less observed data.

The second layer of the hierarchical prior has virtually an identical structure to (\ref{eq:policy_player}) only here the position-specific parameter vector $\boldsymbol \beta^{(g,y,z)}$ is given a multivariate normal prior with mean vector $\boldsymbol \gamma^{(y,z)}$, which denotes the average propensity to shoot in court region $y$ under defensive pressure $z$, regardless of player or position.  For the final stage of the hierarchy, (\ref{eq:policy_location}), we use the 8-dimensional 0-vector as the prior mean, yielding a 0.5 prior probability of shooting in any giving region/defense combination.  While this is an unrealistic shot probability for many states, our model is not sensitive to the values for this prior mean given the immense amount of data we have for each region/defense combination across the season in conjunction with weakly informative priors on each $\Sigma$ in the hierarchy.

Each $\boldsymbol{\Sigma}$ in (\ref{eq:policy_player})-(\ref{eq:policy_location}) is an AR(1) covariance matrix with variance parameters and correlation parameters corresponding to their respective levels of the hierarchy (i.e. $\rho_{\theta},\rho_{\beta}, \rho_{\gamma}$, and $\sigma^2_{\theta},\sigma^2_{\beta}, \sigma^2_{\gamma})$.  We chose this covariance structure under the assumption that player shooting behavior is more similar over intervals of the shot clock that are close in time, and that this correlation diminishes as intervals between times increases.  The AR(1) structure provides a natural construct to model this type of correlation.    Specifically, $\rho$ represents the correlation between shot propensity parameters in the same state of the MDP for adjacent intervals of the shot clock (e.g. (12,15] and (15, 18]).

\begin{figure}[H]
\begin{center}
\includegraphics[width=5in, trim={.3cm .3cm .3cm .3cm}, clip]{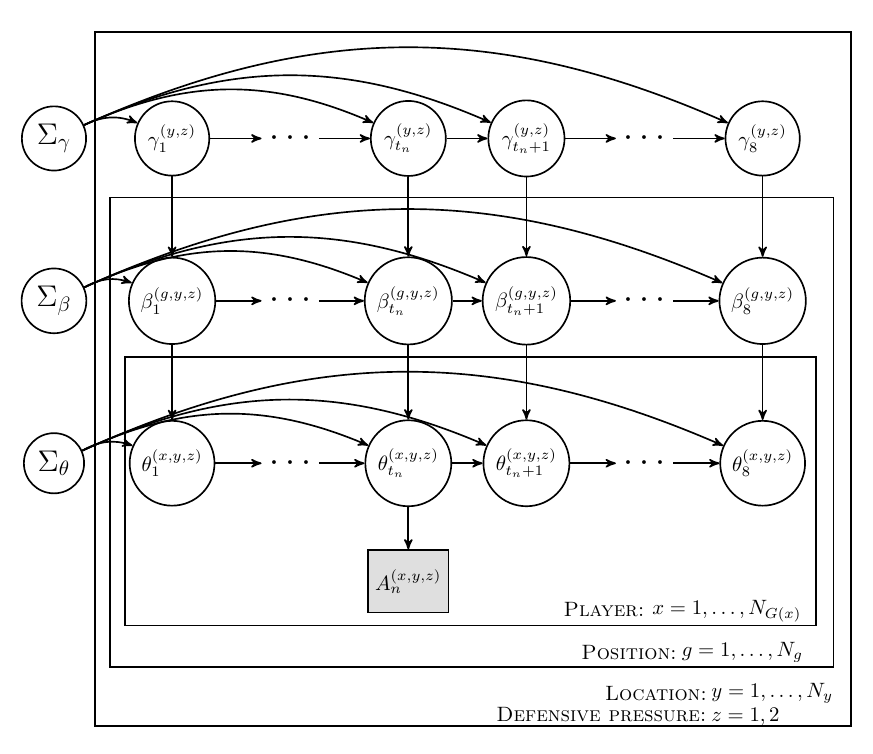} 
\end{center}
\setlength{\belowcaptionskip}{0pt}
\setlength{\abovecaptionskip}{-5pt}
\caption{A graphical representation of our model for $\pi(\cdot)$.  The observable random variable is denoted by a gray box while parameters are denoted by unfilled circles.  $A^{(x,y,z)}_n$ denotes the action of player $x$ in region $y$ with defensive pressure $z$ during time interval $t_n$.  It is governed by the corresponding state's propensity parameter, $\theta^{(x,y,z)}_{t_n}$.  These player/region/defense parameters $\boldsymbol \theta$ are modeled by a multi-stage hierarchical prior, including a layer for position/region/defense parameters $\boldsymbol \beta$, and region/defense parameters $\boldsymbol \gamma$.  The parameters $\boldsymbol \theta$, $\boldsymbol \beta$, and $\boldsymbol \gamma$ all have a temporal dimension of $n_{\scriptscriptstyle{\text{TPT}}}= 8$, which we convey horizontally in each layer of the graph.  These multivariate parameter vectors have latent AR(1) covariance matrices, $\Sigma_{\theta}$, $\Sigma_{\beta}$, $\Sigma_{\gamma}$.  As shown by the plates in the figure, player-specific parameters are nested within position-specific parameters, while location and defensive pressure comprise the base model state space.}
\label{fig:graphical_model}
\end{figure}

We use half-Cauchy priors \citep{polson2012half} on all the scale parameters and Uniform[0, 1) priors on the temporal correlation parameters:
\begin{align*}
\sigma_{\theta}, \sigma_{\beta}, \sigma_{\gamma} \sim \text{half-Cauchy}(0,2.5) ~~~~~~~\rho_{\theta}, \rho_{\beta}, \rho_{\gamma} \sim \text{Uniform}[0,1).
\end{align*}
The $\text{half-Cauchy}(0,2.5)$ prior contains 90\% of its mass in the interval (0, 15.78) which is sufficiently non-informative for our application, and we only consider positive correlation for $\rho$ given that temporal trends in shooting behavior are generally smooth.  Figure \ref{fig:graphical_model} shows a graphical representation of the model for $\pi(\cdot)$.  

If a shot is taken (i.e. if $a_n = \text{`Shoot'}$) then the MDP episode terminates and the reward, $\Up{R}_{n+1}$, is determined by $R(\cdot)$.  Otherwise, $r_{n+1} = 0$ and the next state is determined by $P(\cdot)$.  

\subsection{Transition probability model---$P(\cdot)$} 

Conditional on $a_n = \text{`Not shoot'}$, we model the probability that the play transitions to state $s^{(x',y',z')}$ as a function of the ballcarrier's latent propensity to transition to state $s^{(x',y',z')}$ given his current state $s^{(x,y,z)}_n$ and the current interval of the shot clock $t_n$:
\begin{align}
P(s,a,s') = \mathbb{P}(\Up{S}_{n+1} = s^{(x',y',z')} | a_n, s^{(x,y,z)}_n, t_n, \boldsymbol \lambda) &= \frac{\text{exp}\Big(\lambda^{\big((x,y,z),(x',y',z')\big)}_{t_n}\Big)}{\sum\limits_{i,j,k} \text{exp}\Big(\lambda^{\big((x,y,z),(i,j,k)\big)}_{t_n}\Big)} ,
\end{align}
where $\Up{S}_{n+1}$ is a categorical random variable and $\lambda^{\big((x,y,z),(x',y',z')\big)}_{t_n}$ denotes player $x$'s propensity to transition to $s^{(x',y',z')}$ when in court region $y$ under defensive pressure $z$.  The symbols $a_n$, $s^{(x,y,z)}_n$, and $t_n$ are all as defined in Section \ref{sec:policy_model}.  The support of $\Up{S}_{n+1}$ is equivalent to the state space $\mathcal{S}^{team}$ defined in Section \ref{sec:state_and_action} with one additional state---the terminal state representing a turnover.  Note that $\boldsymbol \lambda$ is a massive 3-dimensional parameter array with dimensions $|\mathcal{S}^{team}| \times (|\mathcal{S}^{team}|+1) \times n_{\scriptscriptstyle{\text{TPT}}}$.  For a 15-player roster using 3-second shot clock intervals for the TPT this results in a $(180 \times 181 \times 8)$ array, yielding a minimum of 260,640 player-specific parameters to estimate for a single team.

As with the models for $\pi(\cdot)$ and $R(\cdot)$, we employ multi-level hierarchical priors for $\boldsymbol \lambda$:
\begin{align}
 \boldsymbol \lambda^{\big((x,y,z),(x',y',z')\big)} &\sim \mathbf{N}_8\Big(\boldsymbol \zeta^{\big((G(x),y,z), (G(x'),y',z')\big)}, \boldsymbol \Sigma_{\lambda}\Big) \label{eq:P_top} \\
\boldsymbol \zeta^{\big((g,y,z), (g,y',z')\big)} &\sim \mathbf{N}_8\Big(\boldsymbol \omega^{\big((y,z),(y',z')\big)}, \boldsymbol \Sigma_{\zeta}\Big), \label{eq:P_stage1} \\ 
 \boldsymbol \omega^{\big((y,z), (y',z')\big)} ~ &\sim \mathbf{N}_8\big(\mathbf{0}, \boldsymbol \Sigma_{\omega}\big). \label{eq:P_stage2}
\end{align}
In (\ref{eq:P_top}), $\boldsymbol \zeta^{\big((g,y,z),(g',y',z')\big)}$ denotes the the 8-dimensional average propensity of all players (i.e. all players $\{x'_1, \ldots, x'_{N_h}\}$ for whom $H(x'_i) = H(x)$) $x$ with $G(x) \equiv g$ in court region $y$ under defensive pressure $z$ to transition to any state with $G(x') \equiv g'$, in court region $y'$ under defensive pressure $z'$ over the 8 discrete intervals of the shot clock. In (\ref{eq:P_stage1}), the position-specific parameter vector $\boldsymbol \zeta^{\big((g,y,z), (g,y',z')\big)}$ is given a multivariate normal prior with mean vector $\boldsymbol \omega^{\big((y,z),(y',z')\big)}$, which can be similarly defined at the global region/defense level.  As in the model for $\pi(\cdot)$, we use the 8-dimensional 0-vector as the prior mean for $\boldsymbol \omega^{\big((y,z),(y',z')\big)}$ and each $\boldsymbol{\Sigma}$ is an AR(1) covariance matrix with variance parameters and correlation parameters corresponding to their respective levels of the hierarchy (i.e. $\rho_{\lambda}, \rho_{\zeta},\rho_{\omega}$, and $\sigma^2_{\lambda}, \sigma^2_{\zeta}, \sigma^2_{\omega}$).

Due to the computational burden of jointly fitting this 3-stage model (i.e. jointly fitting all 30 team's transition dynamics---a total of nearly 10 million parameters), we use a two-stage modeling approach for $P(\cdot)$ which was imperative given our computational constraints.  This enables us to fit each team's transition dynamics \textit{separately} while still allowing us to borrow strength from the position-specific and location/defense levels of the hierarchy.  The details of the two-stage fitting process for $P(\cdot)$ are included in \ref{suppA}.

\subsection{Reward function---$R(\cdot)$}
 
In context of a basketball play, (\ref{eq:R}) can be restated as, ``How many immediate points do we expect when a player in state $s$ takes action $a$?".  If the action is a shot, then this expected value is his expected points per shot from the given state, otherwise it is 0.\footnote{In our analysis we have omitted plays that ended in fouls and any free throw situations.}  This allows us to define the reward function of the MDP completely in terms of a shot efficiency model. 

Prior to formally defining $R(\cdot)$, we propose a model for the probability that a shot is made.  Given a shot, we model the make-probability as a function of the shooter's skill and a region-specific additive effect if the shot was uncontested, or open: 
\begin{align}
 \text{Make}(s) = \mathbb{P}(M_n = 1 | s^{(x,y,z)}_n, \boldsymbol \mu,\boldsymbol \xi) &= \text{expit}\Big(\mu^{(x, y)} + I(z_n = \text{`Open'}) \times \xi^{(y)}\Big),
 \end{align}
where $M_n$ is a Bernoulli random variable for whether the attempted shot in moment $n$ was made, $\mu^{(x,y)}$ denotes player $x$'s contested shooting skill in court region $y$, $I(\cdot)$ is an indicator function of the defensive pressure $z_n$ in moment $n$, and $\xi^{(y)}$ is the effect in court region $y$ if the shot is uncontested.  Note that in this model, defensive pressure is a region-specific additive effect rather than being built into the player-specific parameters.  This is because detecting player-specific differences in how defensive pressure affects their shot-make probability is impracticable.  While minor differences certainly exist, it would take massive amounts of shot data---more than we have---to detect these differences with statistical confidence.   

The reward function $R(\cdot)$ is simply the scaled make-probabilities for each state if shot is taken (scaled by 2 or 3, depending on the court-region), and 0 in the case that a shot is not taken or if a turnover occurs:
\begin{align}
R(s, a) =  
\begin{cases} 
     3 \times \text{Make}(s), &y \in \{\text{3-pointer}\}, a_n = \text{`Shot'}\\
     2 \times \text{Make}(s), & y \in \{\text{2-pointer}\}, a_n = \text{`Shot'}\\
     0, & \text{otherwise}.
  \end{cases} \label{eq:R}
\end{align}

As with $P(\cdot)$ and $\pi(\cdot)$, we use a multi-stage hierarchical prior for the player-specific parameters $\boldsymbol \mu$:  
\begin{align}
\mu^{(x, y)}  &\sim \text{N}(\psi^{(H(x),y)}, \sigma_{\mu}^2) \label{eq:r_prior_1} \\
\psi^{(h,y)}  &\sim \text{N}(\varphi^{y}, \sigma_{\psi}^2)  \label{eq:r_prior_2} \\
\varphi^{y} &\sim \text{N}(0, \sigma_{\varphi}^2).
\end{align}
In (\ref{eq:r_prior_1}), $\mu^{(x, y)}$ is given a normal prior with mean vector $\psi^{(H(x),y)}$, which denotes the average shooting skill of all players who are share the same group as player $x$ (i.e. all players $\{x'_1, \ldots, x'_{N_h}\}$ for whom $H(x'_i) = H(x)$) in location $y$.    We emphasize that while $P(\cdot)$ and $\pi(\cdot)$ define player groups $\mathbf{g}$ using naive player positions (center, power forward, point guard, etc.) this model uses new groups, $\mathbf{h}$, on which to base the regularization.  The reason for this change is that a player's shooting skill does not have as clear a correspondence to his naive position.  As such, we create customized groupings to ensure sensitivity to this variation.  We will detail how we constructed these new groups shortly.

In the second layer of the hierarchical prior, $\psi^{(h,y)}$ is given a normal prior with mean vector $\varphi^{y}$, which denotes the global average shooting skill from court region $y$.  As in the model for $\pi(\cdot)$ and $P(\cdot)$, the final stage of the hierarchy is given a mean-0 normal prior with variance $\sigma_{\varphi}^2$.  Note that all the parameters in this model are given univariate priors rather than multivariate priors since we model a player's shooting skill as being constant over the shot clock.

Next we define the priors for $\boldsymbol \xi$, the region-specific additive effects for contested shots.
\begin{align}
\xi^y &\sim \text{half-normal}(0, \sigma_{\xi}^2) \label{eq:xi_prior} 
\end{align} 
We use a half-normal prior distribution since, all else being equal, open shots have higher make probabilities than contested shots.  Finally, as with the scale hyperpriors for $\pi(\cdot)$, we use half-Cauchy priors for the scale parameters in $R(\cdot)$:
\begin{align*}
\sigma_{\mu}, \sigma_{\psi}, \sigma_{\varphi}, \sigma_{\xi} &\sim \text{half-Cauchy}(0,2.5). 
\end{align*}

\vspace{-.2cm}
To create the new player groupings $\mathbf{h}$, we first clustered players into three categories based exclusively on the volume of shots they took over the course of the season, irrespective of the shot locations.  Next, we re-clustered players into six shot propensity categories based on the proportional breakdown of their shots by court region, irrespective of volume.  In both clustering procedures we used the k-means algorithm initialized at cluster centroids calculated via Ward linkage \citep{ward1963}.  We then crossed these clusters, giving a total of 18 groups which differentiate players by how much they shoot and where they tend to shoot from.  Table \ref{tab:player_table} shows three example players in each cluster.  
\begin{table}[ht]
\centering
\caption{Players were independently clustered by shot volume and shot region propensity.  The table shows three players in each group after crossing the resulting clusters.}
\label{tab:player_table}
\resizebox{\textwidth}{!}{
\begin{tabular}{P{1.5cm}|rrrrrr}
  \hline
  \multirow{2}{*}{\parbox{1.25cm}{Shot Volume}}
      & \multicolumn{6}{c}{Shot Region Propensity}  \\      \cline{2-7}
  & Equal Balance & 3-point Heavy &  Mid Heavy  & Rim Heavy & 3-point Specialist &  Rim Specialist  \\  \hline
\multirow{3}{*}{High}  & L. James & D. Lillard & J. Wall & A. Davis & S. Curry & A. Drummond \\      
 & R. Westbrook & K. Love & D. Nowitzki  & I. Thomas &  T. Ariza & G. Monroe \\    
 & D. Cousins& J. Harden & K. Leonard & D. Wade & W. Matthews & J. Okafor \\      
 \hline
\multirow{3}{*}{Med}  & L. Barbosa  &  P. Beverly & R. Rubio &  D. Favors & K. Korver  &  D. Jordan \\    
 & L. Stephenson &  E. Ilyasova & M. Speights & E. Turner  &  J. Terry  & T. Booker  \\     
 & N. Jokic & O. Porter & M. Belinelli & B. Portis & N. Mirotic  & C. Capela  \\    
 \hline
\multirow{3}{*}{Low}  & A. Roberson & J. Jerebko & M. Muscala &  A. Miller & J. Ingles & A. Bogut  \\      
 & D. Motiejunas & B. Jennings & C. Watson & A. Varejao & J. Ennis & B. Bass \\    
 & K. McDaniels & D. Augustin & T. Prince & D. Powell & B. Rush & J. McGee \\    
\hline
\end{tabular}}
\end{table}

We also assume independence across plays for shot make probabilites, which is a debated area of research (\cite{neiman2011reinforcement} for example). 

\subsection{Inference and validation}

After removing plays we are not interested in modeling (plays that terminated in either fouls, timeouts, jumpballs, or in the backcourt) we have 155,656 plays ($\approx1.93$ million observations) on which we fit our models.  We held out a sample of approximately 28,000 plays to use for model validation.  We fit our models using Stan, an open-source software package which offers a suite of MCMC methods for statistical inference \citep{carpenter2017stan}.  For each model we initialized two chains and let them mix long enough to ensure we had a potential scale reduction factor $< 1.05$ for every parameter.  Effective sample sizes ranged from 48 to 15,000 across the set of parameters (see \ref{suppA} for additional diagnostics).  Details on the Stan model scripts for $\pi(\cdot)$, $P(\cdot)$, and $R(\cdot)$ can be found in the companion GitHub repository for this paper.

Following \cite{franks2015characterizing} and \cite{cervone2016multiresolution}, both of which utilize Bayesian hierarchical models in conjunction with NBA optical tracking data, we use out-of-sample log-likelihood as a mechanism for model validation.  Table \ref{tab:Model_validation} shows out-of-sample log-likelihoods for four models of increasing complexity for each component of the MDP.  The transition model column, $P(\cdot)$, represents log-likelihoods computed using only the Cleveland Cavaliers TPT model, whereas the other columns comprise the entire league.  For all three components, the models with player-specific shrinkage perform best.  All subsequent references to MDP model components refer to the models in row D of Table \ref{tab:Model_validation}.   
\begin{table}[h]
\begin{center}
\caption{Out-of-sample log-likelihoods for four models of increasing complexity over each component of the MDP.}
\label{tab:Model_validation}
\begin{tabular}{lccc}
\toprule
Model & $\pi(\cdot)$ & $P(\cdot)$ & $R(\cdot)$\\
 \hline
\text{A. Empirical model} & -36808  & -17299  & -5956  \\ 
\text{B. Model A + location shrinkage} & -25187  & {-38702}  & {-4571} \\
\text{C. Model B + position shrinkage}  & {-24467} & -25099  & {-4561} \\
\text{D. Model C + player shrinkage} & {\textbf{-21553}}  & \textbf{-13478} & {\textbf{-4540}} \\
\bottomrule
\end{tabular}
\end{center}
\end{table}

\subsection{Model fit}
\begin{figure}[H]
\begin{center}
\includegraphics[width=5in]{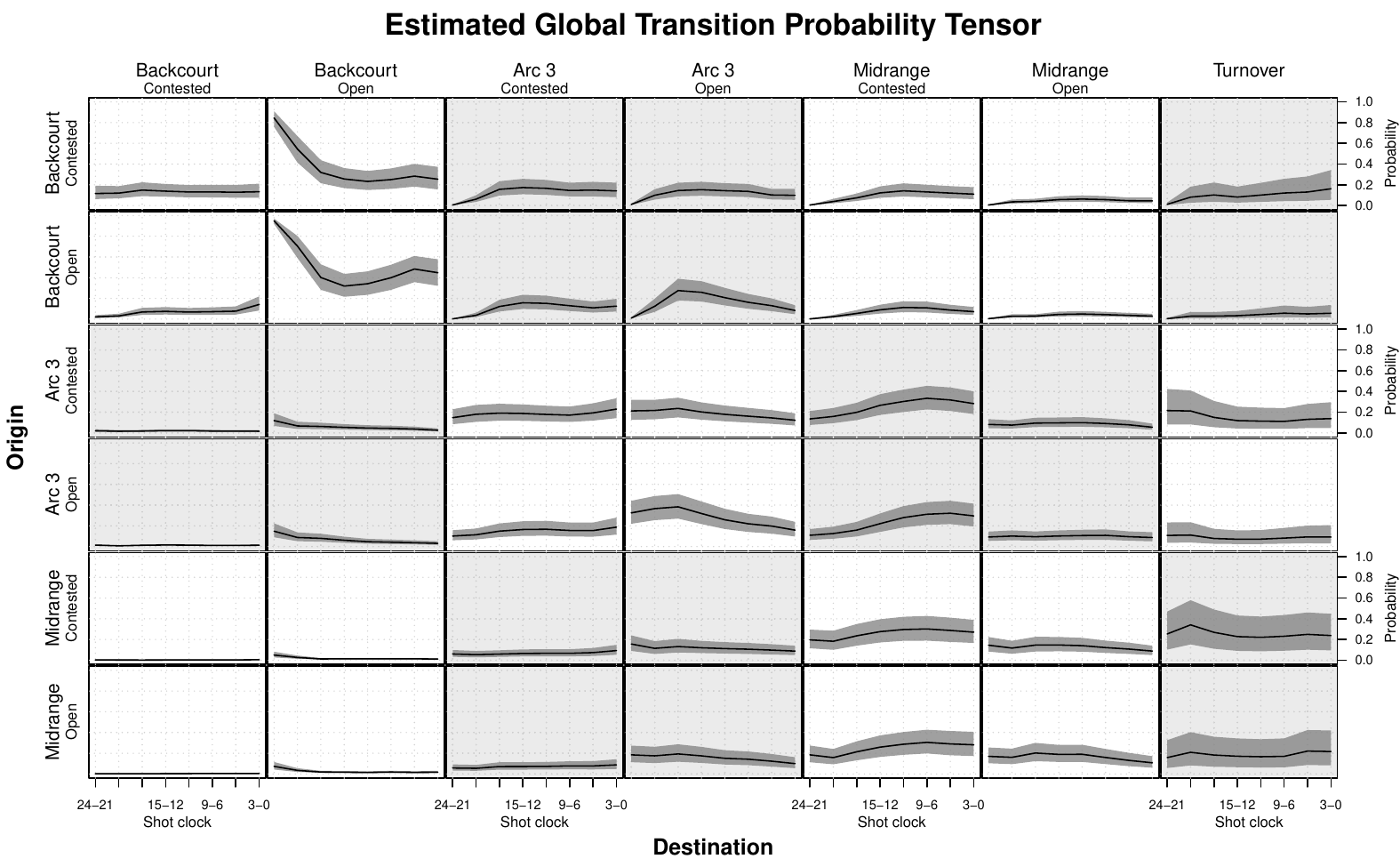}
\end{center}
\setlength{\belowcaptionskip}{0pt}
\setlength{\abovecaptionskip}{-5pt}
\caption{Estimated league-wide transition probability tensor for the top level of the hierarchy on which each team's TPT is built.  Within each plot frame, the 95\% credible interval of the origin to destination transition probability is shown in dark gray and the posterior mean is shown by a black line. Within each plot frame the x-axis represents time on the shot clock, while the y-axis represents the transition probability.  Across plot frames, the y-axis represents the origin state and the x-axis represents the destination state. Corner 3, paint, and rim states are omitted to maintain a practical size for the figure.}
\label{fig:Global_TPT}
\end{figure}

Figure \ref{fig:Global_TPT} shows 95\% credible intervals for the transition probabilities in the top hierarchy of the TPT model.  
As shown in the block diagonal frames of the figure,  the highest transition probabilities are to the same state, due to the predominant influence of dribbles in the data.  Conversely, it is improbable for the ball to transition immediately to a state which is not directly geographically adjacent.  Interestingly, the defensive pressure of the destination state appears to have a much larger impact on transition probabilities than the defensive pressure of the origin state.  

\begin{figure}[H]
\begin{center}
\includegraphics[width=5in, trim={0cm 0cm 0cm 1cm}, clip]{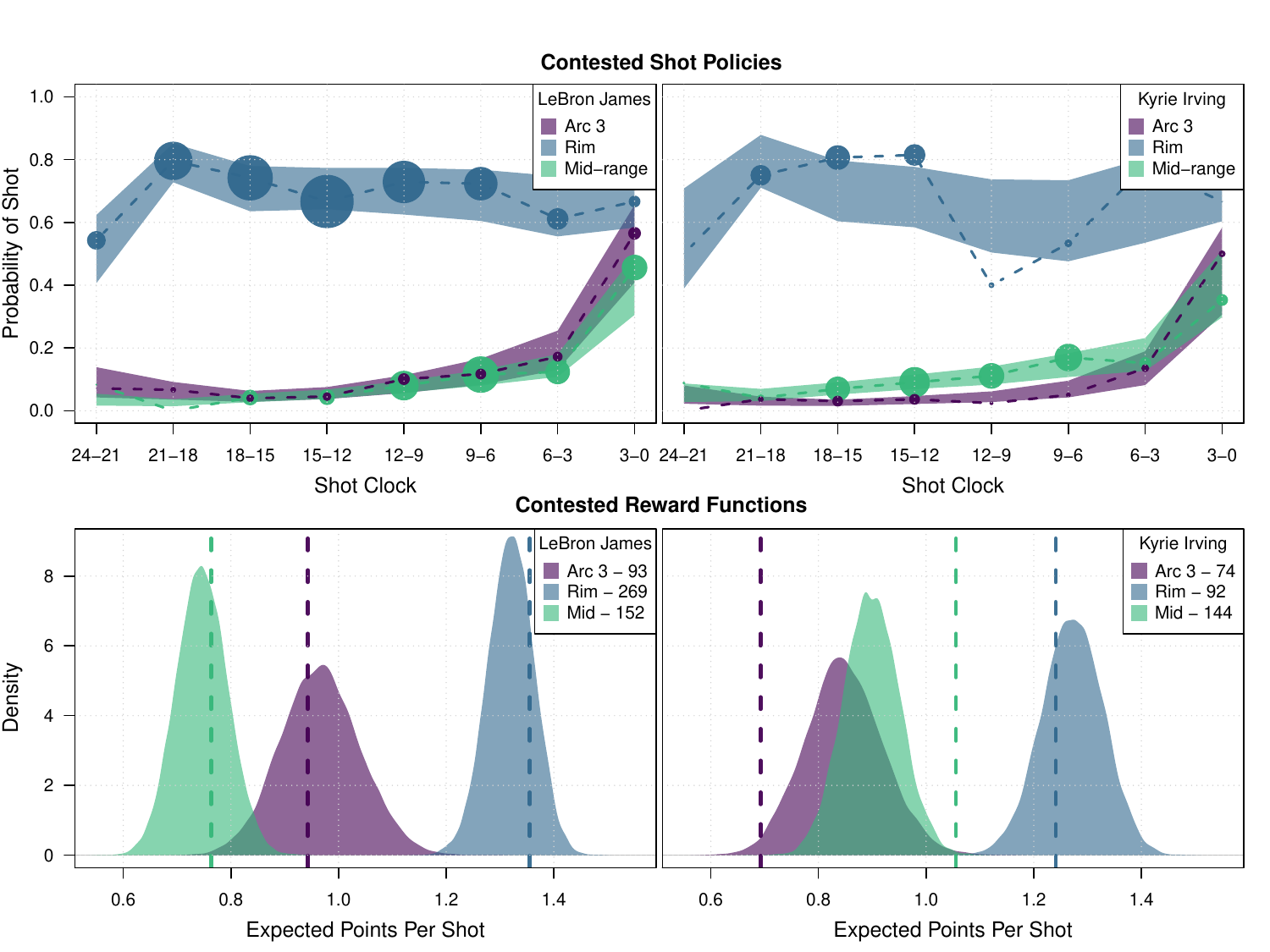}
\end{center}
\setlength{\belowcaptionskip}{-5pt}
\setlength{\abovecaptionskip}{0pt}
\caption{Estimated shot policies (95\% credible intervals) and reward functions (posterior densities) for LeBron James and Kyrie Irving in three sample states.  The shot policies are overlaid with dots corresponding to their empirical shot policies; the sizes of the dots are relative to how many shots they took within that time interval from the indicated state.  The reward functions are also overlaid with the empirical points per shot and the number of shots they took from each state is given in the legend.}
\label{fig:joint_policy_rewards}
\end{figure}

The estimated shot policies and reward functions for LeBron James and Kyrie Irving of the Cleveland Cavaliers are shown in Figure \ref{fig:joint_policy_rewards}.  The strong temporal autocorrelation captured by the model ($\widehat{\rho}_{\theta} = 0.94$) significantly smooths jagged empirical policies yielding more plausible results.  The two players' policies look quite similar, with the exception that Irving tends to take contested mid-range shots more frequently than James.  
Interestingly, Irving also takes contested mid-range shots more frequently than he takes arc 3-point shots.  In general, this is considered poor shot selection because most players have a higher expected points per shot (EPPS) from beyond the arc than from the mid-range.  However, Irving appears to be an anomaly in this respect;  his mid-range reward distribution is greater than his arc 3 distribution in expectation.  His estimated shot policy evidences that he knows his strengths and acts accordingly.  

\section{Simulating plays}
\label{sec:sim}

Having fit the models for the latent components of the MDP, our next task is to simulate plays using these models.  In this section, we first describe our algorithm for simulating a basketball play and we conclude the section by comparing our simulations to the observed trajectories. 
 
\subsection{Play simulation algorithm} The algorithm requires seven inputs: $s_0$, $\boldsymbol \theta$, $\boldsymbol \lambda$,  $\boldsymbol \mu$, $\boldsymbol \xi$, $c_{0}$, and $L(\cdot)$.  The first piece, $s_0$, denotes the starting state of a play.  For these we use the observed starting states for each team's collection of plays in the 2015-2016 NBA regular season.  Note that we do not treat the number of plays in a season nor the states in which plays begin as random.  Consequently, we do not analyze rebounding; the number of plays is fixed beforehand and once a turnover happens or a shot is taken, the play ends.

Next we require parameters to govern the components of the MDP which stochastically generate the states visited in plays, when shots occur, and the point values of taken shots.  For these inputs we use the model fits for these components described in Section \ref{sec:tensor_model_spec} (i.e. $\boldsymbol \theta$ for $\pi(\cdot)$, $\boldsymbol \lambda$ for $P(\cdot)$, and ($\boldsymbol \mu$, $\boldsymbol \xi$) for $R(\cdot)$).  Specifically, we take a random draw from each parameter's respective set of Markov chains used to approximate its posterior distribution, $\widehat{f}(\cdot | \mathcal{D})$, where $\mathcal{D}$ denotes the training data.  By using random draws from $\widehat{f}(\cdot | \mathcal{D})$ for each parameter rather than a single functional of the estimated posterior distribution (e.g. the posterior means), the uncertainty in our estimation gets propagated through to our simulations.  In other words, a range of plausible parameter values are used in the simulations, rather than a single point estimates.  This is critical in accurately quantifying the uncertainty in our simulations.  We denote a posterior draw from an arbitrary parameter's posterior distribution by a tilde over the parameter (e.g. $\widetilde{\boldsymbol \lambda}$ denotes a posterior draw from $\widehat{f}(\boldsymbol \lambda | \mathcal{D})$). 

Lastly, we must account for the shot clock, including starting shot clock times for all plays and a mechanism to take time off the shot clock at each step of the MDP.   As in the case for $s_0$, we use the observed shot clock times at the start of each play, denoted $c_0$, for each team's collection of plays in the 2015-2016 NBA regular season.  To take time off of the shot clock at each step of the MDP, we sample the team's empirical distribution of time-lapses between events conditional on the current interval of the shot clock.  This component of the simulator makes performing analytical operations on the process intractable because the distribution of time lapses between events does not lend itself to a parametric distribution.  We denote this empirical distribution by ${L}(\cdot)$, which is a function of $t_n$.  Algorithm 1 details the conceptual structure of the simulation process.
\begin{figure}[H]
\begin{algorithm}[H]
\SetAlgoLined
 \KwData{$s_0$, $\widetilde{\boldsymbol \theta}$, $\widetilde{\boldsymbol \lambda}$,  $\widetilde{\boldsymbol \mu}$, $\widetilde{\boldsymbol \xi}$, $c_{0}$, $L(\cdot)$}
 \KwResult{List of the simulated states (terminal and intermediary), actions, and rewards}
 n $\leftarrow$ 0\;
 \While{$s_n \neq \text{Turnover}$}{
   $t_{n} \leftarrow T(c_n)$ \;
  $a_{n} \leftarrow$ Bernoulli variate from $\pi\big(\cdot | \widetilde{\boldsymbol \theta}, s_n, t_n\big)$ \;
    \uIf{$a_{n} = $ Shot}{
       $r_{n+1} \leftarrow R\big(s_n, a_n | \widetilde{\boldsymbol \mu}, \widetilde{\boldsymbol \xi}\big)$ \;
      \textbf{break loop}\;
   }
   \Else{
      $r_{n+1} \leftarrow 0$\;
  }
  lapse $\leftarrow$ draw from $L(t_{n})$ \;
  $c_{n+1} \leftarrow c_{n}$ - lapse \;
      \uIf{$c_{n+1} < 0$}{
    $s_{n+1} \leftarrow Turnover$ (shot clock violation) \;
    $a_{n+1} \leftarrow $ NULL \;
    $r_{n+1} \leftarrow 0$ \;
    }
     \Else{
     $s_{n+1} \leftarrow$ categorical variate from $P\big(\cdot | \widetilde{\boldsymbol \lambda}, s_n, t_n\big)$ \; 
  }
    $ n \leftarrow n + 1$ \;
 }
 \Return{$\{\boldsymbol s, \boldsymbol a, \boldsymbol r\}$}
 \caption{Basketball play simulator}
\end{algorithm}
\end{figure}

In Algorithm 1, $n$ indexes the sequential events in the play.  The while-loop iteratively generates actions, states, and 0-valued rewards until either 1) a shot is taken, at which point the reward is determined by $R(\cdot)$ and the loop breaks (i.e. the play terminates), or 2) the play transitions to the `Turnover' state, which can also occur by the shot clock expiring.  For each play we keep track of all generated states, actions, and rewards.

\subsection{Calibration}

We can be extremely detailed in checking the calibration of our simulations since we keep track of all simulated intermediary and terminal transitions. To assess the calibration, we simulate 300 seasons for the Cleveland Cavaliers using the observed starting states of all their 2015-16 plays and compare our simulations to the actual transition counts.  Note that these simulations are on-policy, meaning they are computed using variates of the shot policy estimated on the observed data.   In making this comparison we must be cognizant of overfitting; the empirical model will always yield optimal calibration because the empirical model fits both trends and errors.  Models with regularization may appear less calibrated, but ultimately give better fits because the modeling of errors is attenuated by the induced shrinkage.  

\begin{figure}[ht]
\hbox{\hspace{-2em} 
\includegraphics[trim={0cm .25cm 0cm 1cm},clip,width=5.25in]{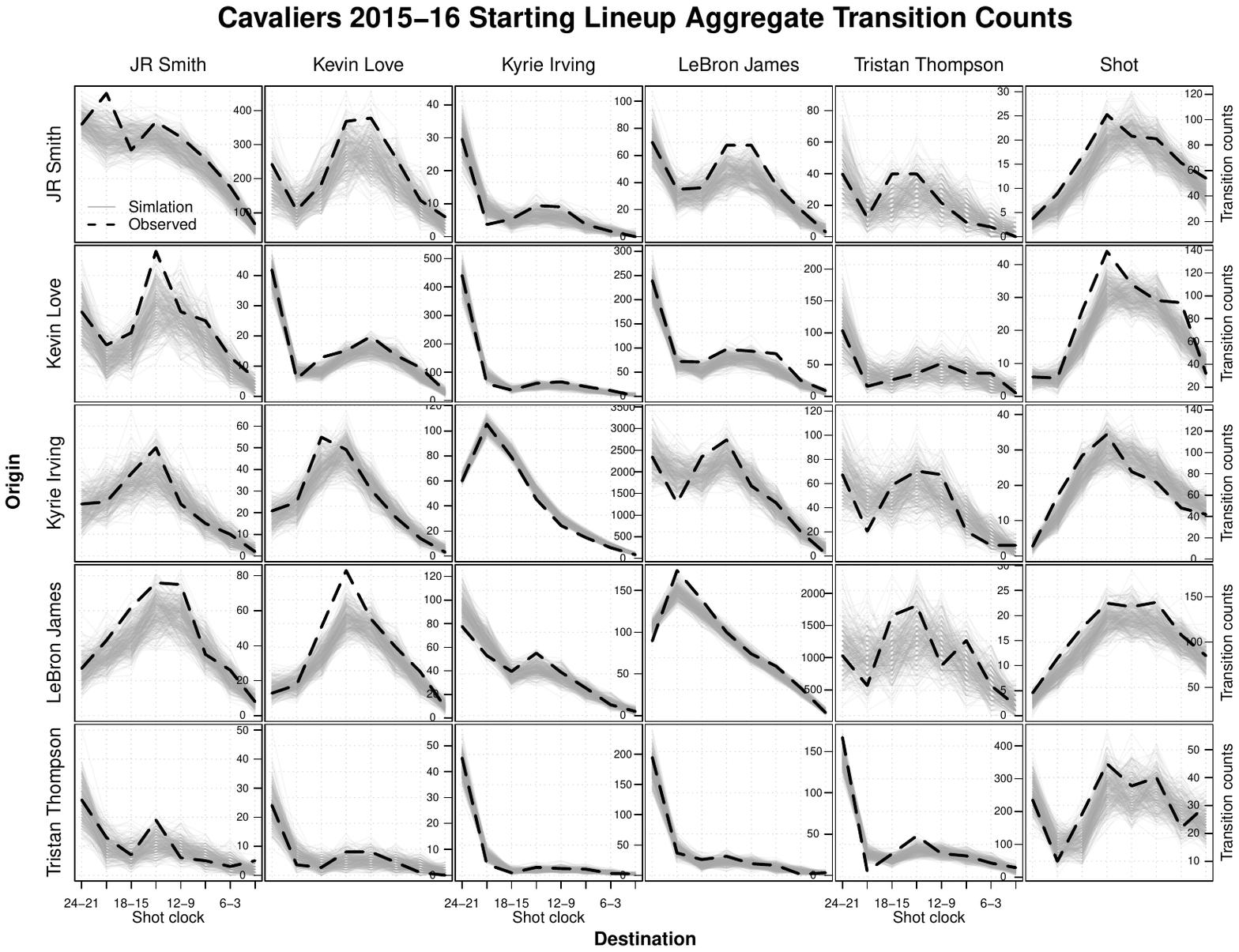}}
\setlength{\belowcaptionskip}{-5pt}
\setlength{\abovecaptionskip}{5pt}
\caption{300 simulated (gray) season-total transition counts over the shot clock overlaid with the corresponding observed counts (black) for the 2015-2016 season.  Within each plot frame, the x-axis represents time on the shot clock, while the y-axis represents total transition counts.  Across plot frames, the y-axis represents the origin state and the x-axis represents the destination state.}   
\label{fig:calibration}
\end{figure}

Figure \ref{fig:calibration} shows the simulated player-aggregate transition counts for these 300 simulations for the Cavaliers' starting lineup overlaid with the observed counts in red.  Our simulations capture the aggregate transition count trends over the shot clock with high integrity; however, they appear to be slightly biased low for some state pairs.  On the other hand, simulated transition counts for low-usage players (not shown) are generally biased high.  As noted previously, these phenomena are due to shrinkage in the hierarchical model, which we are quick to note is not a model deficiency.  As evidenced in Table \ref{tab:Model_validation}, this borrowing of information improves out-of-sample model fit, giving us more reliable calibration on macro-level features.  

\section{Altering policies}

With confidence that the method accurately reproduces play sequences under the observed policy model, we now simulate team-specific plays under altered shot policies.  However, before providing examples of altered policies we pause to discuss some relevant topics from game-theory.  

\subsection{Game theory}

\subsubsection{Optimal policies}  A general assumption of this paper is that teams are not operating on optimal shot policies.  This is difficult to test but there is research that supports this conclusion \citep{goldman2014optimal, skinner2012}.  Regarding optimal stopping times (i.e. when a player shoots during a play relative to the shot clock), \cite{goldman2014optimal} show that while on average, the league as a whole closely follows the optimal curve, individual players are not perfect optimizers, often exhibiting tendency to undershoot.  Even under the assumption that a team is operating optimally, players and coaches could still gain utility by exploring adverse effects of changes to this policy.  

\subsubsection{Allocative efficiency}  A player's shot efficiency depends on the volume of opportunities he is allocated.  The mathematical formulation of this concept originates in the work of \cite{oliver2004}.  Oliver defines the relationship between a player's usage and his efficiency as a ``skill curve" and suggests that it should generally exhibit a downward trend, meaning that players become less efficient as they carry more of the scoring load.  This relationship is important in context of altering shot policies.  As explained in \cite{goldman2014misperception}, if a team changes its shot policy to take more 3-point shots, the team has to accept lower quality 3-point opportunities on the margin.  This will lead to lower expected values for these additional shots but higher expected values for the 2-point shots that by consequence have a lower usage rate due to the increase in 3's.  There is a counter-balancing relationship for policy changes due to moving up (or down) the skill curve.  For our purposes, the simulation method should not bias the results of testing policy changes, as long as the changes are not drastic.  

\subsubsection{Defensive response}  If a team makes a tactical change that gives them an advantage, it is reasonable to assume that the defense will attempt to eliminate the advantage.  This defensive response brings up some important questions in context of our project --- ``How sensitive are defenses to policy alterations?" and ``How long does it take for a defense to respond sufficiently to make a policy change ineffectual?"  These questions depend on too many variables to suggest a single answer; however, we offer some observational evidence from the past two NBA regular seasons that suggests that, in some cases, the defensive response resulting from a team's altered shot policy does not render its strategy change ineffectual over the course of a season.    

In the 2016-17 NBA regular season, the Toronto Raptors shot on average 30.5\% of their shots from 3-point range and they averaged 1.1 points per shot on these attempts.\footnote{These statistics were gathered from stats.nba.com.}  In the  2017-18 season, they shot 39.6\% of their shots from 3-point range.  This represents a 30\% increase in their team 3-point shot policy.  Despite this  increase in their 3-point shot policy, the Raptors' expected points per shot (EPPS) from beyond the arc only decreased less than 2\% (from 1.1 in 2016-17 to 1.08 in 2017-18).  Additionally, the Raptors' overall EPPS increased from 1.1 in 2016-17 to 1.14 in 2017-2018.  So while the policy change resulted in a small loss of efficiency (perhaps due to defensive adaptation), the response was not such that it rendered the Raptors' policy change a zero-sum net benefit.  

We acknowledge that this example is observational; these results could be due to season-to-season variability or the outcome of other variables, such as the addition of new players or the development/decline of returning players.  Ultimately, predicting season outcomes for alternate policies is an extrapolation.  As such, we believe that testing minor perturbations to a team's policy will yield more credible results and that proposed changes should be carefully crafted prior to testing. 

\subsection{Shot policy changes}

We implement on-policy simulations by turning the crank of Algorithm 1 using a team's 2015-16 collection of starting state and shot clock time pairs ($s_0, c_0$) for all of their regular season plays in tandem with posterior draws from the MDP model fits.  This results in one simulation of a season.  Simulating a season with a policy alteration follows the same process with one additional step---for each simulation we transform the posterior draw of the shot policy model according to our alteration specifications, then simulate seasons with these modified posterior draws.  An example may help clarify exactly how we perform this computationally.
 
Suppose we want to test a policy change in which a specific player $x_i$ shoots in all court regions and at all intervals of the shot clock with a 10\% increase in frequency.  We first modify the posterior draws of the policy parameters $\widetilde{\boldsymbol \theta}$ according to the specified alteration for all affected elements:
\begin{align}
\widetilde{\theta}^{(x,y,z)^{alt}}_{t_n} = (0.1 \times \widetilde{\theta}^{(x,y,z)}_{t_n}) + \widetilde{\theta}^{(x,y,z)}_{t_n} ~~&\forall ~~ y, z, t_n,\text{ and } x \text{ such that }x \equiv x_i,
\end{align}
where $\widetilde{\theta}^{(x,y,z)^{alt}}_{t_n}$ represents an altered element of $\widetilde{\boldsymbol \theta}$ and all other symbols are as defined previously in the text.\footnote{We set a maximum threshold for altered shot policy parameters.  If any of the parameters that would be altered end up with $\widetilde{\theta}^{(x,y,z)^{alt}}_{t_n} > 0.9$, we cap the alteration at 0.9.  For reasonable policy alterations this issue shouldn't arise.}   

\begin{figure}[h]
\begin{center}
\includegraphics[width=5in]{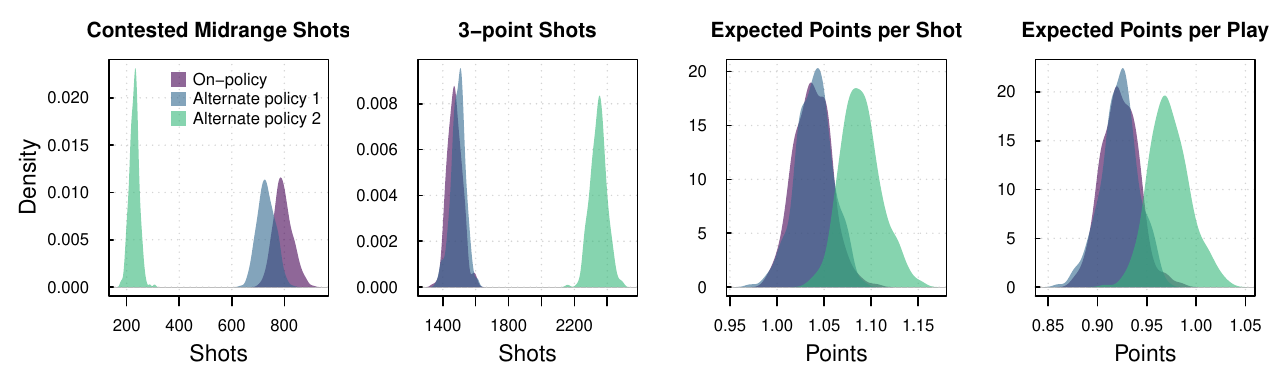}
\end{center}
\setlength{\belowcaptionskip}{-5pt}
\setlength{\abovecaptionskip}{0pt}
\caption{Left to right: distribution of simulated contested mid-range shots, 3-point shots, expected points per shot, and expected points per play.}
\label{fig:altered_shot_policies}
\end{figure}

We now show two examples of policy changes that could be explored with our methods and compare the altered policy simulations to on-policy simulations.   For each policy, we simulate 300 seasons for the Cleveland Cavaliers.  The results are shown in Figure \ref{fig:altered_shot_policies}.  

\begin{quote}
\textbf{Alteration 1.}  Reduce the contested mid-range shot policy by 20\% for all players on the team while more than 10 seconds remain on the shot clock.  

\textbf{Alteration 2.}  Regardless of time on the shot clock, reduce all contested mid-range shot policies by 70\% while doubling all three-point shot policies.  
\end{quote}

The most obvious distinction between the policies is the divergence between the contested mid-range and 3-point shot distributions, which is not surprising since we directly altered these shot policies.  However, in order to measure whether the policy yields a net positive result for a given team, we must quantify how the altered policy affects efficiency and production.   To measure these effects, we restrict our attention to the differences in EPPS and expected points per play (EPPP). Under policy 2, shot efficiency increases ({1.038} to {1.089} in EPPS) as does play production ({0.923} to {0.973} in EPPP).  Under policy 1, these distributions show no practical differences, largely due to only 7.5\% of plays ending in a mid-range shot with over 10 seconds on the shot clock, limiting the potential impact.  

\subsection{Passing policy changes}
\label{sec:passing_changes}

\begin{figure}[h]
\begin{center}
\includegraphics[width=5in]{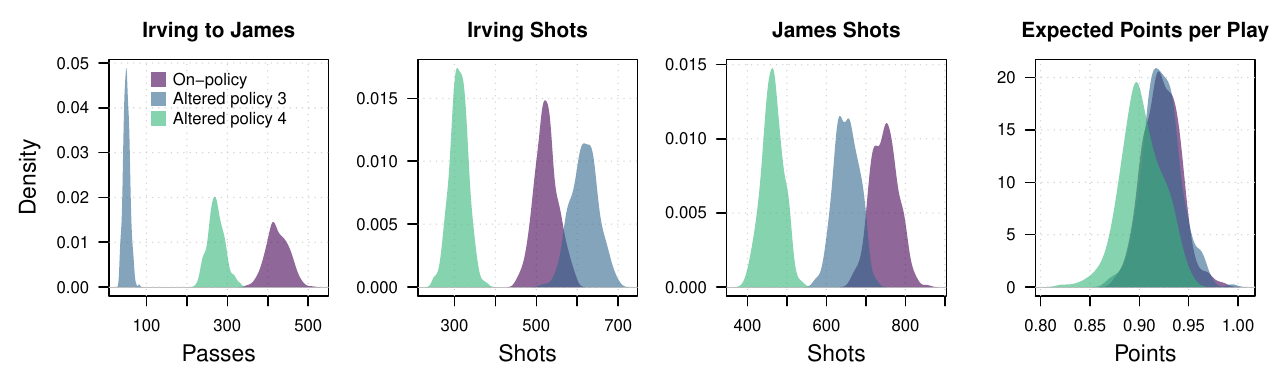}
\end{center}
\setlength{\belowcaptionskip}{-5pt}
\setlength{\abovecaptionskip}{0pt}
\caption{Left to right: distributions of simulated transitions from Irving to James, Irving's total shots, James' total shots, and expected points per play.}
\label{fig:selfish_kyrie}
\end{figure}

With a few modifications we can consider broader policy changes that encompass not only shooting but passing and movement as well.  This entails altering the probabilities of non-terminating state transitions via the TPT.\footnote{In addition to the game theoretic consequences mentioned previously, new complexities arise with altering passing/movement policies in the context of our model framework.  Many state-transition pairs in the TPT are not physically possible (e.g. a player cannot transition directly from the backcourt to the restricted area).  Also, any change where we \textit{increase} player-to-player transition probabilities is potentially problematic.  Passing \textit{more often} to a player in a specific location hinges on the assumption that the other player is correspondingly available in that location, which is something we do not control in our model.}  We now explore two altered policies of this nature; the results are shown in Figure \ref{fig:selfish_kyrie}.

\begin{quote}
\textbf{Alteration 3.}  Reduce the transitions from Irving to James by 90\%.  

\textbf{Alteration 4.}  Triple the transition probabilities from all veterans to players on rookie contracts, while reducing the transition probabilities from rookie contract players to veterans by 75\%.  
\end{quote}

Alteration 3 represents a pathological example in which Irving forces his way into being the dominant player on the team by almost never distributing the ball to James.  The downstream effects of the dominant Irving policy lead to a {18\%} increase his expected total shot count, while James' is reduced by {13\%}.  Interestingly, though Irving's and James' total shot distributions change dramatically, our method predicts that the overall differences in production would be negligible.  

Alteration 4 could be described as a youth development policy, where veteran players are asked to take a back seat and players on rookie contracts are given the green light on offense.  This policy change has a much larger predicted impact on production.  We estimate this policy change would cost the Cavaliers {0.02} EPPP, which could have significant consequences on win totals and playoff outcomes.  

\section{Conclusion}

We have developed and implemented a method to test the impact of shot-clock dependent policy adjustments over the course of a season at an unprecedented level of detail while accounting for model uncertainty in every aspect of the system.  These methods could have immediate practical impact across multiple levels of a basketball organization.  Coaches could assess proposed strategy changes outside of games rather than risking poor results by testing them in games.  Front offices could explore the performance of hypothetical rosters by leveraging the position-level transition probabilities in tandem with their player-specific shot policies and reward functions.  These tools could prove useful in evaluating trades and in free-agency negotiations.   Additionally, our methods could enable teams to gauge the effects of having to play second-string players if any starters suffered a long-term injury.  The examples we have shown in this paper are only the tip of the iceberg in terms of how these methods could be utilized.    

We have primarily considered shooting decisions in this introductory work, but as shown in Section \ref{sec:passing_changes}, our methodology naturally scales to include all different types of basketball decisions, allowing coaches and analysts to explore incredibly nuanced tactical changes.  Additionally, with similar tracking data now available for most major sports including hockey, football, and soccer, our methods could extend to testing decision policies in other sports.

In a broader statistical context, we have provided and implemented a novel framework for modeling within-episode non-stationarity in MDPs through the use of policy and transition probability tensors.  We have also shown how to combine multiple MDPs into a single weighted average process, which can enable solutions to problems that were previously impracticable to compute.  Additionally, we've built a method to simulate from this type of MDP when the arrival times cannot be modeled parametrically.  These contributions could be beneficial in many different areas such as traffic modeling, queuing applications, and environmental processes.  

Our paper opens doors for promising further studies.  In terms of reinforcement learning, a clear next step would be to solve/estimate the action-value function for the functional MDP we introduce in this paper.  In the basketball context, addressing the game theoretic aspects by incorporating usage curves and simulating defensive response could make these methods more robust.

\begin{appendix}

\section{Additional details and diagnostics}

\subsection{MDP in context of a basketball play}
$~$
 \begin{figure}[H]
\begin{center}
\includegraphics[width=5in]{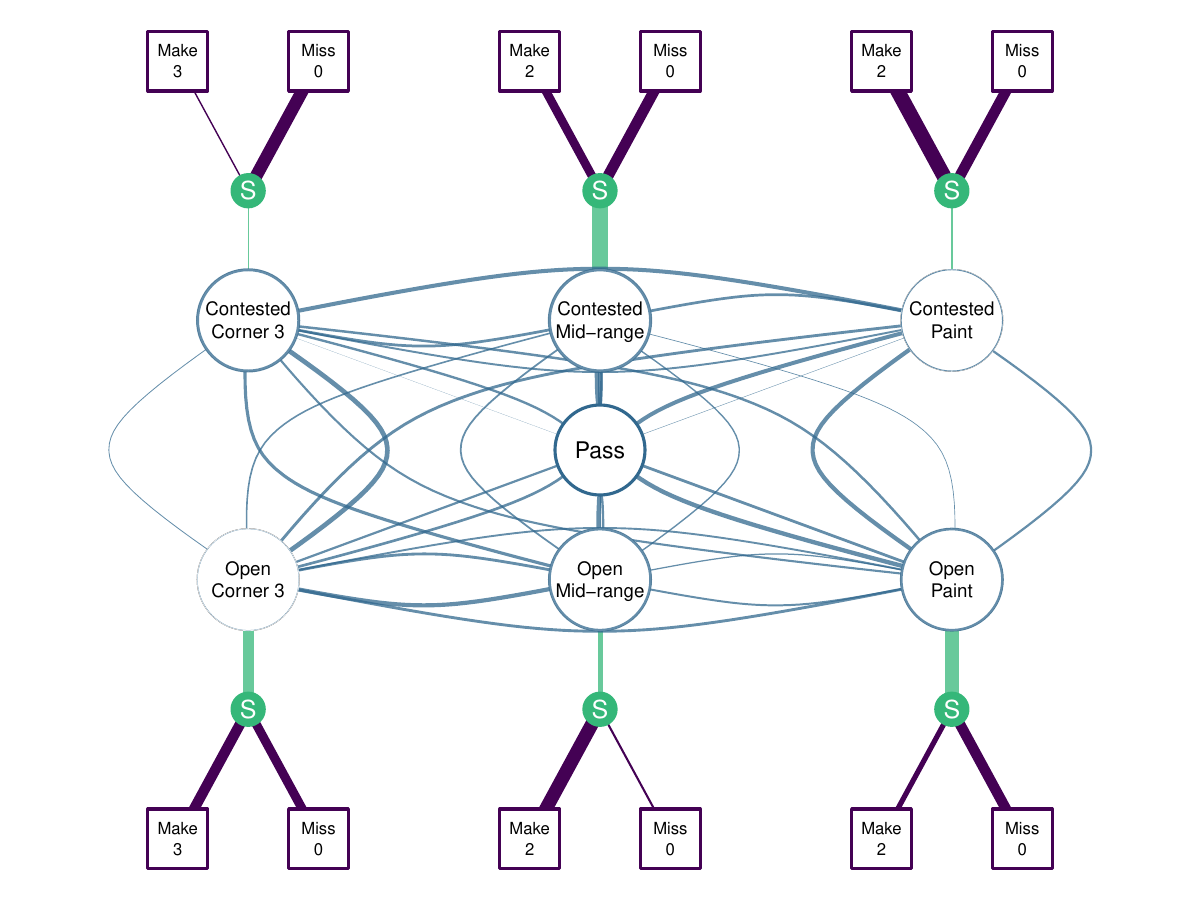} 
\end{center}
\caption{Illustration of the components of the MDP for a single player in context of a basketball play. The blue circles represent states, the solid green circles represent actions (shots), and the curved blue lines represent transition probabilities between states.  The green lines of varying width connecting the blue state circles to the green action circles represent the policy.  The purple lines connecting the green action circles to the squares represent the reward function.   Players may pass the ball to another player (not shown) which is also a transition to a non-terminal state.}
\label{fig:graph_pic}
\end{figure}

\newpage
\subsection{MCMC diagnostics}
$~$

\begin{table}[H]
\begin{center}
\caption{MCMC details and diagnostics for each fitted model.  $P_1(\cdot)$ and $P_2(\cdot)$ refer to the 1st and 2nd stages of $P(\cdot)$.  Specifically, $P_2(\cdot)$ refers to the 2nd stage fit on the Cavaliers TPT.}
\label{tab:Model_diagnostics}
\begin{tabular}{lcccc}
\toprule
Parameter & $\pi(\cdot)$ & $P_1(\cdot)$ & $P_2(\cdot)$ & $R(\cdot)$\\
 \hline
\text{MCMC samples (per chain)} & 2000 &16000 & 1500 &  11000\\
\text{Burn-in} & 500 & 1000 & 500 & 1000 \\
\text{Minimum eff. sample size}  & {48} & {76} & {371} & {181}\\
\text{Maximum} $\hat{R}$ & {1.039} & {1.022} & {1.000} & {1.026}\\
\bottomrule
\end{tabular}
\end{center}
\end{table}

\subsection{Defensive pressure rules} 
\begin{align}
\Up{S}^{defense}_n =  
\begin{cases} 
    \text{contested}, & s^{region}_n \equiv \text{rim } \& \text{ ndd} < 3 \\
    \text{contested}, & s^{region}_n \equiv \text{paint } \& \text{ ndd} < 3.5 \\
    \text{contested}, & s^{region}_n \equiv \text{mid-range } \& \text{ ndd} < 4 \\
    \text{contested}, & s^{region}_n \in \{\text{corner 3 }, \text{arc 3 }, \text{back-court}\}~  \& \text{ ndd} < 5 \\
    \text{open}, & \text{otherwise.}
\end{cases} \label{eq:s_defense}
\end{align}

\section{Two-stage approximation for $P(\cdot)$}

This section describes our two stage approximation for the transition probability function $P(\cdot)$.  We first lay out both stages in mathematical terms, then define all terms and the mechanics of the fitting process. 
\begin{center}
\textbf{STAGE 1}
\end{center}
\begin{align}
\boldsymbol \zeta^{\big((g,y,z), (g,y',z')\big)} &\sim \mathbf{N}_8\Big(\boldsymbol \omega^{\big((y,z),(y',z')\big)}, \boldsymbol \Sigma_{\zeta}\Big), \label{eq:P_stage1} \\ 
 \boldsymbol \omega^{\big((y,z), (y',z')\big)} ~ &\sim \mathbf{N}_8\big(\mathbf{0}, \boldsymbol \Sigma_{\omega}\big), \label{eq:P_stage2}
\end{align}
\begin{center}
\textbf{STAGE 2}
\end{center}
\begin{align}
 \boldsymbol \lambda^{\big((x,y,z),(x',y',z')\big)} &\sim \mathbf{N}_8\Big(\widehat{\boldsymbol \zeta}^{\big((G(x),y,z), (G(x'),y',z')\big)}, \mathbf{S}_{\lambda}^{\text{approx.}}\Big)
\end{align}
In stage 1, all symbols are as defined in the main text for equations (3.7) and (3.8).  We fit these two layers of $P(\cdot)$ exactly as described for $\pi(\cdot)$ in Section 3.1---the hyperpriors for $\sigma_{\zeta}$ and $\sigma_{\omega}$ are half-Cauchy(0, 2.5) and the corresponding correlation parameters are Uniform[0,1).

After fitting stage 1, we can fit stage 2 separately for each team by using a fixed covariance matrix $\mathbf{S}_{\lambda}^{\text{approx.}}$ in the multivariate normal prior on $\boldsymbol \lambda$.  For a given team, we initialize the prior means of the player-specific parameters, $\boldsymbol \lambda $, at the corresponding posterior means of the position-specific parameter estimates, $\widehat{\boldsymbol \zeta}$, estimated in stage 1.

Choosing the prior variance for $\boldsymbol \lambda$ requires greater discretion, since the covariance matrix governs how strongly $\boldsymbol \lambda$ will shrink toward each player's position-specific parameter estimate $\widehat{\boldsymbol \zeta}$ from stage 1.  As in our model for $\pi(\cdot)$, $\mathbf{S}_{\lambda}^{\text{approx.}}$ is given an AR(1) structure, therefore fixing the covariance matrix amounts to selecting values for $\sigma_{\lambda}$ and $\rho_{\lambda}$.  We assume that the temporal covariance structure governing the shot policy model $\pi(\cdot)$ resembles that of the transition model $P(\cdot)$.  This allows us to leverage our knowledge about of the ratio between $\sigma_{\theta}$ and $\sigma_{\beta}$ in selecting a value for $\sigma_{\lambda}$.  Using the posterior means as estimates, $\Big(\frac{\widehat{\sigma}_{\theta}}{\widehat{\sigma}_{\beta}}\Big) = 1.8$, hence the player-specific parameters have a standard deviation roughly twice as big as that of the group-specific parameters in the policy model.  Since we have an estimate for the group-level standard deviation of the transition model from stage 1, $\widehat{\sigma}_{\zeta}$, we multiply this value by the ratio from the policy model, 1.8, to select an appropriate value for $\widehat{\sigma}_{\lambda}$.  Hence we fix $\sigma_{\lambda} = \widehat{\sigma}_{\zeta} \times 1.8 = 2.14$.  Likewise, we fix $\rho_{\lambda}$ at the corresponding value in the policy model, which is 0.94.  We can then create the covariance matrix $\mathbf{S}_{\lambda}^{\text{approx.}}$ and fit stage 2 of the approximation to get estimates for $\boldsymbol \lambda$.

\end{appendix}

 \section*{Acknowledgements}
This work was partially supported by NSERC.
 

\bibliographystyle{agsm}
\bibliography{aoas_refs}

\end{document}